\documentclass{esub2acm}

\newcommand{\lsb}{\mbox{$[\! [$}} \newcommand{\rsb}{\mbox{$]\! ]$}}

\def\picture #1 by #2 (#3){ \vbox to #2{ \hrule width #1 height 0pt
depth 0pt \vfill \special{picture #3} } }

\def\scaledpicture #1 by #2 (#3 scaled #4){{ \dimen0=#1 \dimen1=#2
\divide\dimen0 by 1000 \multiply\dimen0 by #4 \divide\dimen1 by 1000
\multiply\dimen1 by #4 \picture \dimen0 by \dimen1 (#3 scaled #4)} }

\newcommand{\comment}[1]{}

\begin{document}

\title{Machine Learning in Automated Text Categorization}

\author{Fabrizio Sebastiani \\ Consiglio Nazionale delle Ricerche,
Italy}

\begin{abstract}

The automated categorization (or classification) of texts into
predefined categories has witnessed a booming interest in the last ten
years, due to the increased availability of documents in digital form
and the ensuing need to organize them. In the research community the
dominant approach to this problem is based on machine learning
techniques: a general inductive process automatically builds a
classifier by learning, from a set of preclassified documents, the
characteristics of the categories. The advantages of this approach
over the knowledge engineering approach (consisting in the manual
definition of a classifier by domain experts) are a very good
effectiveness, considerable savings in terms of expert manpower, and
straightforward portability to different domains. This survey
discusses the main approaches to text categorization that fall within
the machine learning paradigm. We will discuss in detail issues
pertaining to three different problems, namely document
representation, classifier construction, and classifier evaluation.

\end{abstract}







\begin{bottomstuff}

\begin{authinfo} \address{Istituto di Elaborazione dell'Informazione,
Consiglio Nazionale delle Ricerche, Via G.\ Moruzzi, 1, 56124 Pisa
(Italy). E-mail: {\tt fabrizio@iei.pi.cnr.it}}
\end{authinfo}

\end{bottomstuff}

\markboth{F.\ Sebastiani} {Machine Learning in Automated Text
Categorization}

\maketitle


 
\section{Introduction} \label{sec:intro}

In the last ten years content-based document management tasks
(collectively known as \emph{information retrieval} -- IR) have gained
a prominent status in the information systems field, due to the
increased availability of documents in digital form and the ensuing
need to access them in flexible ways. \emph{Text categorization} (TC
-- aka \emph{text classification}, or \emph{topic spotting}), the
activity of labelling natural language texts with thematic categories
from a predefined set, is one such task. TC dates back to the early
'60s, but only in the early '90s it became a major subfield of the
information systems discipline, thanks to increased applicative
interest and to the availability of more powerful hardware. TC is now
being applied in many contexts, ranging from document indexing based
on a controlled vocabulary, to document filtering, automated metadata
generation, word sense disambiguation, population of hierarchical
catalogues of Web resources, and in general any application requiring
document organization or selective and adaptive document dispatching.

Until the late '80s the most popular approach to TC, at least in the
``operational'' (i.e.\ real-world applications) community, was a {\em
knowledge engineering} (KE) one, consisting in manually defining a set
of rules encoding expert knowledge on how to classify documents under
the given categories. In the '90s this approach has increasingly lost
popularity (especially in the research community) in favour of the
\emph{machine learning} (ML) paradigm, according to which a general
inductive process automatically builds an automatic text classifier by
learning, from a set of preclassified documents, the characteristics
of the categories of interest. The advantages of this approach are an
accuracy comparable to that achieved by human experts, and a
considerable savings in terms of expert manpower, since no
intervention from either knowledge engineers or domain experts is
needed for the construction of the classifier or for its porting to a
different set of categories. It is the ML approach to TC that this
paper concentrates on.

Current-day TC is thus a discipline at the crossroads of ML and IR,
and as such it shares a number of characteristics with other tasks
such as \emph{information/knowledge extraction from texts} and {\em
text mining} \cite{Knight99,Pazienza97}. There is still considerable
debate on where the exact border between these disciplines lies, and
the terminology is still evolving. ``Text mining'' is increasingly
being used to denote all the tasks that, by analyzing large quantities
of text and detecting usage patterns, try to extract probably useful
(although only probably correct) information. According to this view,
TC is an instance of text mining. TC enjoys quite a rich literature
now, but this is still fairly scattered\footnote{A fully searchable
bibliography on TC created and maintained by this author is available
at {\tt
http://liinwww.ira.uka.de/bibliography/Ai/automated.text.categorization.html}}.
Although two international journals have devoted special issues to
this topic \cite{Joachims01,Lewis94b}, there are no systematic
treatments of the subject: there are neither textbooks nor journals
entirely devoted to TC yet, and \cite[Chapter16]{Manning99} is the
only chapter-length treatment of the subject. As a note, we should
warn the reader that the term ``automatic text classification'' has
sometimes been used in the literature to mean things quite different
from the ones discussed here. Aside from (i) the automatic assignment
of documents to a predefined set of categories, which is the main
topic of this paper, the term has also been used to mean (ii) the
automatic identification of such a set of categories (e.g.\
\cite{Borko63}), or (iii) the automatic identification of such a set
of categories {\em and} the grouping of documents under them (e.g.\
\cite{Merkl98}), a task usually called \emph{text clustering}, or (iv)
any activity of placing text items into groups, a task that has thus
both TC and text clustering as particular instances \cite{Manning99}.

This paper is organized as follows. In Section \ref{sec:definition} we
formally define TC and its various subcases, and in Section
\ref{sec:applications} we review its most important applications.
Section \ref{sec:MLincategorization} describes the main ideas
underlying the ML approach to classification. Our discussion of
\emph{text} classification starts in Section \ref{sec:indexingandDR}
by introducing \emph{text indexing}, i.e.\ the transformation of
textual documents into a form that can be interpreted by a
classifier-building algorithm and by the classifier eventually built
by it. Section \ref{sec:construction} tackles the inductive
construction of a text classifier from a ``training'' set of
preclassified documents. Section \ref{sec:evaluation} discusses the
evaluation of text classifiers. Section \ref{sec:conclusion}
concludes, discussing open issues and possible avenues of further
research for TC.


\section{Text categorization} \label{sec:definition}


\subsection{A definition of text categorization}
\label{sec:definitionTC}

Text categorization is the task of assigning a Boolean value to each
pair $\langle d_j,c_i \rangle \in {\cal D} \times {\cal C}$, where
${\cal D}$ is a domain of documents and ${\cal C}=\{c_1, \ldots,
c_{|{\cal C}|}\}$ is a set of predefined \emph{categories}. A value of
$T$ assigned to $\langle d_j,c_i \rangle$ indicates a decision to file
$d_j$ under $c_i$, while a value of $F$ indicates a decision not to
file $d_j$ under $c_i$. More formally, the task is to approximate the
unknown \emph{target function} $\breve{\Phi}: {\cal D}\times {\cal
C}\rightarrow\{T,F\}$ (that describes how documents ought to be
classified) by means of a function $\Phi: {\cal D}\times {\cal
C}\rightarrow \{T,F\}$ called the \emph{classifier} (aka \emph{rule},
or \emph{hypothesis}, or \emph{model}) such that $\breve{\Phi}$ and
$\Phi$ ``coincide as much as possible''. How to precisely define and
measure this coincidence (called \emph{effectiveness}) will be
discussed in Section \ref{sec:effectivenessmeasures}. From now on we
will assume that:

\begin{itemize}

\item The categories are just symbolic labels, and no additional
knowledge (of a procedural or declarative nature) of their meaning is
available.

\item No \emph{exogenous} knowledge (i.e.\ data provided for
classification purposes by an external source) is available;
therefore, classification must be accomplished on the basis of
\emph{endogenous} knowledge only (i.e.\ knowledge extracted from the
documents). In particular, this means that \emph{metadata} such as
e.g.\ publication date, document type, publication source, etc.\ is
not assumed to be available.

\end{itemize}

\noindent The TC methods we will discuss are thus completely general,
and do not depend on the availability of special-purpose resources
that might be unavailable or costly to develop. Of course, these
assumptions need not be verified in operational settings, where it is
legitimate to use any source of information that might be available or
deemed worth developing \cite{Diaz98,Junker97}. Relying only on
endogenous knowledge means classifying a document based solely on its
semantics, and given that the semantics of a document is a {\em
subjective} notion, it follows that the membership of a document in a
category (pretty much as the relevance of a document to an information
need in IR \cite{Saracevic75}) cannot be decided deterministically.
This is exemplified by the phenomenon of \emph{inter-indexer
inconsistency} \cite{Cleverdon84}: when two human experts decide
whether to classify document $d_j$ under category $c_i$, they may
disagree, and this in fact happens with relatively high frequency. A
news article on Clinton attending Dizzy Gillespie's funeral could be
filed under {\tt Politics}, or under {\tt Jazz}, or under both, or
even under neither, depending on the subjective judgment of the
expert.


\subsection{Single-label vs.\ multi-label text categorization}
\label{sec:SLCandMLC}

Different constraints may be enforced on the TC task, depending on the
application. For instance we might need that, for a given integer $k$,
exactly $k$ (or $\leq k$, or $\geq k$) elements of ${\cal C}$ be
assigned to each $d_{j}\in {\cal D}$. The case in which exactly 1
category must be assigned to each $d_{j}\in {\cal D}$\comment{(as
e.g.\ in
\cite{Baker98,Cohen98,Guthrie94,Koller97,Joachims97,Larkey99,Li98a,Moulinier96a,Schutze95})}
is often called the \emph{single-label} (aka \emph{non-overlapping
categories}) case, while the case in which any number of categories
from $0$ to $|{\cal C}|$ may be assigned to the same $d_{j}\in {\cal
D}$ is dubbed the \emph{multi-label} (aka \emph{overlapping
categories}) case. A special case of single-label TC is \emph{binary}
TC, in which each $d_{j}\in {\cal D}$ must be assigned either to
category $c_i$ or to its complement $\overline{c}_i$.

From a theoretical point of view, the binary case (hence, the
single-label case too) is more general than the multi-label, since an
algorithm for binary classification can also be used for multi-label
classification: one needs only transform the problem of multi-label
classification under $\{c_1,\ldots, c_{|{\cal C}|}\}$ into $|{\cal
C}|$ independent problems of binary classification under
$\{c_i,\overline{c}_i\}$, for $i=1, \dots, |{\cal C}|$. However, this
requires that categories are stochastically independent of each other,
i.e.\ that for any $c', c''$ the value of $\breve{\Phi}(d_j,c')$ does
not depend on the value of $\breve{\Phi}(d_j,c'')$ and viceversa; this
is usually assumed to be the case (applications in which this is not
the case are discussed in Section \ref{sec:Yahoo}). The converse is
not true: an algorithm for multi-label classification cannot be used
for either binary or single-label classification. In fact, given a
document $d_j$ to classify, (i) the classifier might attribute $k>1$
categories to $d_j$, and it might not be obvious how to choose a
``most appropriate'' category from them; or (ii) the classifier might
attribute to $d_j$ no category at all, and it might not be obvious how
to choose a ``least inappropriate'' category from ${\cal C}$.

In the rest of the paper, unless explicitly mentioned, we will deal
with the binary case. There are various reasons for this:

\begin{itemize}

\item The binary case is important in itself because important TC
applications, including filtering (see Section \ref{sec:filtering}),
consist of binary classification problems (e.g.\ deciding whether
$d_{j}$ is about {\sf Jazz} or not). In TC, most binary classification
problems feature unevenly populated categories (e.g.\ much fewer
documents are about {\sf Jazz} than are not) and unevenly
characterized categories (e.g.\ what is about {\sf Jazz} can be
characterized much better than what is not).

\item Solving the binary case also means solving the multi-label case,
which is also representative of important TC applications, including
automated indexing for Boolean systems (see Section
\ref{sec:indexingforBoolean}).

\item Most of the TC literature is couched in terms of the binary
case.

\item Most techniques for binary classification are just special cases
of existing techniques for the single-label case, and are simpler to
illustrate than these latter.

\end{itemize}

\noindent This ultimately means that we will view classification under
${\cal C}=\{c_1,\ldots, c_{|{\cal C}|}\}$ as consisting of $|{\cal
C}|$ independent problems of classifying the documents in ${\cal D}$
under a given category $c_i$, for $i=1, \dots, |{\cal C}|$. A {\em
classifier for $c_i$} is then a function $\Phi_i:{\cal
D}\rightarrow\{T,F\}$ that approximates an unknown target function
$\breve{\Phi}_i:{\cal D}\rightarrow\{T,F\}$.


\subsection{Category-pivoted vs.\ document-pivoted text
categorization} \label{sec:CPCandDPC}

There are two different ways of using a text classifier. Given
$d_{j}\in{\cal D}$, we might want to find all the $c_{i}\in{\cal C}$
under which it should be filed (\emph{document-pivoted categorization}
-- DPC); alternatively, given $c_{i}\in{\cal C}$, we might want to
find all the $d_{j}\in{\cal D}$ that should be filed under it ({\em
category-pivoted categorization} -- CPC). This distinction is more
pragmatic than conceptual, but is important since the sets ${\cal C}$
and ${\cal D}$ might not be available in their entirety right from the
start. It is also relevant to the choice of the classifier-building
method, as some of these methods (see e.g.\ Section
\ref{sec:examplebased}) allow the construction of classifiers with a
definite slant towards one or the other style.

DPC is thus suitable when documents become available at different
moments in time, e.g.\ in filtering e-mail. CPC is instead suitable
when (i) a new category $c_{|{\cal C}|+1}$ may be added to an existing
set ${\cal C}=\{c_1, \ldots, c_{|{\cal C}|}\}$ after a number of
documents have already been classified under ${\cal C}$, \emph{and}
(ii) these documents need to be reconsidered for classification under
$c_{|{\cal C}|+1}$ (e.g.\ \cite{Larkey99}). DPC is used more often
than CPC, as the former situation is more common than the latter.
 
Although some specific techniques apply to one style and not to the
other (e.g.\ the proportional thresholding method discussed in Section
\ref{sec:thresholding} applies only to CPC), this is more the
exception than the rule: most of the techniques we will discuss allow
the construction of classifiers capable of working in either mode.


\subsection{``Hard'' categorization vs.\ ranking categorization}
\label{sec:catandrank}

While a complete automation of the TC task requires a $T$ or $F$
decision for each pair $\langle d_j,c_i\rangle$, a partial automation
of this process might have different requirements.

For instance, given $d_j\in{\cal D}$ a system might simply \emph{rank}
the categories in ${\cal C}=\{c_1, \ldots, c_{|{\cal C}|}\}$ according
to their estimated appropriateness to $d_j$, without taking any
``hard'' decision on any of them. Such a ranked list would be of great
help to a human expert in charge of taking the final categorization
decision, since she could thus restrict the choice to the category (or
categories) at the top of the list, rather than having to examine the
entire set. Alternatively, given $c_i\in{\cal C}$ a system might
simply rank the documents in ${\cal D}$ according to their estimated
appropriateness to $c_i$; symmetrically, for classification under
$c_i$ a human expert would just examine the top-ranked documents
instead than the entire document set. These two modalities are
sometimes called \emph{category-ranking TC} and \emph{document-ranking
TC} \cite{Yang99a}, respectively, and are the obvious counterparts of
DPC and CPC.

Semi-automated, ``interactive'' classification systems \cite{Larkey96}
are useful especially in critical applications in which the
effectiveness of a fully automated system may be expected to be
significantly lower than that of a human expert. This may be the case
when the quality of the training data (see Section
\ref{sec:MLincategorization}) is low, or when the training documents
cannot be trusted to be a representative sample of the unseen
documents that are to come, so that the results of a completely
automatic classifier could not be trusted completely.

In the rest of the paper, unless explicitly mentioned, we will deal
with ``hard'' classification; however, many of the algorithms we will
discuss naturally lend themselves to ranking TC too (more details on
this in Section \ref{sec:thresholding}).


\section{Applications of text categorization}
\label{sec:applications}

TC goes back to Maron's \citeyear{Maron61} seminal work on
probabilistic text classification. Since then, it has been used for a
number of different applications, of which we here briefly review the
most important ones. Note that the borders between the different
classes of applications listed here are fuzzy and somehow artificial,
and some of these may be considered special cases of others. Other
applications we do not explicitly discuss are speech categorization by
means of a combination of speech recognition and TC
\cite{Myers00,Schapire00}, multimedia document categorization through
the analysis of textual captions \cite{Sable00}, author identification
for literary texts of unknown or disputed authorship \cite{Forsyth99},
language identification for texts of unknown language \cite{Cavnar94},
automated identification of text genre \cite{Kessler97}, and automated
essay grading \cite{Larkey98}.


\subsection{Automatic indexing for Boolean information retrieval
systems} \label{sec:indexingforBoolean}

The application that has spawned most of the early research in the
field \cite{Borko63,Field75,Gray71,Heaps73,Maron61}, is that of
automatic document indexing for IR systems relying on a controlled
dictionary, the most prominent example of which is that of Boolean
systems. In these latter each document is assigned one or more
keywords or keyphrases describing its content, where these keywords
and keyphrases belong to a finite set called \emph{controlled
dictionary}, often consisting of a thematic hierarchical thesaurus
(e.g.\ the NASA thesaurus for the aerospace discipline, or the MESH
thesaurus for medicine). Usually, this assignment is done by trained
human indexers, and is thus a costly activity.

If the entries in the controlled vocabulary are viewed as categories,
text indexing is an instance of TC, and may thus be addressed by the
automatic techniques described in this paper. Recalling Section
\ref{sec:SLCandMLC}, note that this application may typically require
that $k_1 \leq x \leq k_2$ keywords are assigned to each document, for
given $k_1,k_2$. Document-pivoted TC is probably the best option, so
that new documents may be classified as they become available. Various
text classifiers explicitly conceived for document indexing have been
described in the literature; see e.g.\
\cite{Fuhr84,Robertson84,Tzeras93}.

Automatic indexing with controlled dictionaries is closely related to
\emph{automated metadata generation}. In digital libraries one is
usually interested in tagging documents by metadata that describe them
under a variety of aspects (e.g.\ creation date, document type or
format, availability, etc.). Some of these metadata are
\emph{thematic}, i.e.\ their role is to describe the semantics of the
document by means of bibliographic codes, keywords or keyphrases. The
generation of these metadata may thus be viewed as a problem of
document indexing with controlled dictionary, and thus tackled by
means of TC techniques.


\subsection{Document organization}
\label{sec:docorganization}

Indexing with a controlled vocabulary is an instance of the general
problem of document base organization. In general, many other issues
pertaining to document organization and filing, be it for purposes of
personal organization or structuring of a corporate document base, may
be addressed by TC techniques. For instance, at the offices of a
newspaper incoming ``classified'' ads must be, prior to publication,
categorized under categories such as {\tt Personals}, {\tt Cars for
Sale}, {\tt Real Estate}, etc. Newspapers dealing with a high volume
of classified ads would benefit from an automatic system that chooses
the most suitable category for a given ad. Other possible applications
are the organization of patents into categories for making their
search easier \cite{Larkey99}, the automatic filing of newspaper
articles under the appropriate sections (e.g.\ {\tt Politics}, {\tt
Home News}, {\tt Lifestyles}, etc.), or the automatic grouping of
conference papers into sessions.


\subsection{Text filtering} \label{sec:filtering}

\emph{Text filtering} is the activity of classifying a stream of
incoming documents dispatched in an asynchronous way by an information
producer to an information consumer \cite{Belkin92}. A typical case is
a newsfeed, where the producer is a news agency and the consumer is a
newspaper \cite{Hayes90a}. In this case the filtering system should
block the delivery of the documents the consumer is likely not
interested in (e.g.\ all news not concerning sports, in the case of a
sports newspaper). Filtering can be seen as a case of single-label TC,
i.e.\ the classification of incoming documents in two disjoint
categories, the relevant and the irrelevant. Additionally, a filtering
system may also further classify the documents deemed relevant to the
consumer into thematic categories; in the example above, all articles
about sports should be further classified according e.g.\ to which
sport they deal with, so as to allow journalists specialized in
individual sports to access only documents of prospective interest for
them. Similarly, an e-mail filter might be trained to discard ``junk''
mail \cite{Androutsopoulos00,Drucker99} and further classify non-junk
mail into topical categories of interest to the user.

A filtering system may be installed at the producer end, in which case
it must route the documents to the interested consumers only, or at
the consumer end, in which case it must block the delivery of
documents deemed uninteresting to the consumer. In the former case the
system builds and updates a ``profile'' for each consumer
\cite{Liddy94}, while in the latter case (which is the more common,
and to which we will refer in the rest of this section) a single
profile is needed.

A profile may be initially specified by the user, thereby resembling a
standing IR query, and is updated by the system by using feedback
information provided (either implicitly or explicitly) by the user on
the relevance or non-relevance of the delivered messages. In the TREC
community \cite{Lewis95b} this is called \emph{adaptive filtering},
while the case in which no user-specified profile is available is
called either \emph{routing} or \emph{batch filtering}, depending on
whether documents have to be ranked in decreasing order of estimated
relevance or just accepted/rejected. Batch filtering thus coincides
with single-label TC under $|{\cal C}|=2$ categories; since this
latter is a completely general TC task some authors
\cite{Hull94,Hull96,Schapire98,Schutze95}, somewhat confusingly, use
the term ``filtering'' in place of the more appropriate term
``categorization''.

In information science document filtering has a tradition dating back
to the '60s, when, addressed by systems of various degrees of
automation and dealing with the multi-consumer case discussed above,
it was called \emph{selective dissemination of information} or {\em
current awareness} (see e.g.\ \cite[Chapter 6]{Korfhage97}). The
explosion in the availability of digital information has boosted the
importance of such systems, which are nowadays being used in contexts
such as the creation of personalized Web newspapers, junk e-mail
blocking, and Usenet news selection.

Information filtering by ML techniques is widely discussed in the
literature: see e.g.\ \cite{Amati99,Iyer00,Kim00,Tauritz00,Yu98}.


\subsection{Word sense disambiguation} \label{sec:WSD}

\emph{Word sense disambiguation} (WSD) is the activity of finding,
given the occurrence in a text of an ambiguous (i.e.\ polysemous or
homonymous) word, the sense of this particular word occurrence. For
instance, {\tt bank} may have (at least) two different senses in
English, as in {\tt the Bank of England} (a financial institution) or
{\tt the bank of river Thames} (a hydraulic engineering artifact). It
is thus a WSD task to decide which of the above senses the occurrence
of {\tt bank} in {\tt Last week I borrowed some money from the bank}
has. WSD is very important for many applications, including natural
language processing, and indexing documents by word senses rather than
by words for IR purposes. WSD may be seen as a TC task (see e.g
\cite{Gale93,Escudero00}) once we view word occurrence contexts as
documents and word senses as categories. Quite obviously, this is a
single-label TC case, and one in which document-pivoted TC is usually
the right choice.

WSD is just an example of the more general issue of resolving natural
language ambiguities, one of the most important problems in
computational linguistics. Other examples, which may all be tackled by
means of TC techniques along the lines discussed for WSD, are {\em
context-sensitive spelling correction}, \emph{prepositional phrase
attachment}, \emph{part of speech tagging}, and \emph{word choice
selection} in machine translation; see \cite{Roth98} for an
introduction.


\subsection{Hierarchical categorization of Web pages}
\label{sec:Yahoo}

TC has recently aroused a lot of interest also for its possible
application to automatically classifying Web pages, or sites, under
the hierarchical catalogues hosted by popular Internet portals. When
Web documents are catalogued in this way, rather than issuing a query
to a general-purpose Web search engine a searcher may find it easier
to first navigate in the hierarchy of categories and then restrict her
search to a particular category of interest.

Classifying Web pages automatically has obvious advantages, since the
manual categorization of a large enough subset of the Web is
infeasible. Unlike in the previous applications, it is typically the
case that each category must be populated by a set of $k_1\leq x \leq
k_2$ documents. CPC should be chosen so as to allow new categories to
be added and obsolete ones to be deleted.

With respect to previously discussed TC applications, automatic Web
page categorization has two essential peculiarities:

\begin{enumerate}

\item \emph{The hypertextual nature of the documents}: links are a
rich source of information, as they may be understood as stating the
relevance of the linked page to the linking page. Techniques
exploiting this intuition in a TC context have been presented in
\cite{Attardi98,Chakrabarti98b,Furnkranz99,Goevert99,Oh00} and
experimentally compared in \cite{Yang01}.

\item \emph{The hierarchical structure of the category set}: this may
be used e.g.\ by decomposing the classification problem into a number
of smaller classification problems, each corresponding to a branching
decision at an internal node. Techniques exploiting this intuition in
a TC context have been presented in
\cite{Dumais00,Chakrabarti98c,Koller97,McCallum98b,Ruiz99,Weigend99}.

\end{enumerate}


\section{The machine learning approach to text categorization}
\label{sec:MLincategorization}

In the '80s the most popular approach (at least in operational
settings) for the creation of automatic document classifiers consisted
in manually building, by means of \emph{knowledge engineering} (KE)
techniques, an expert system capable of taking TC decisions. Such an
expert system would typically consist of a set of manually defined
logical rules, one per category, of type

\begin{center}\textbf{if} $\langle${\sl DNF formula}$\rangle$
\textbf{then} $\langle${\sl category}$\rangle$ \end{center}

\noindent A DNF (``disjunctive normal form'') formula is a disjunction
of conjunctive clauses; the document is classified under $\langle${\sl
category}$\rangle$ iff it satisfies the formula, i.e.\ iff it
satisfies at least one of the clauses. The most famous example of this
approach is the \textsc{Construe} system \cite{Hayes90a}, built by
Carnegie Group for the Reuters news agency. A sample rule of the type
used in \textsc{Construe} is illustrated in Figure \ref{fig:construe}.

\begin{figure} \begin{center}\begin{tabular}{|lcl|} \hline\textbf{if}
& ((\emph{wheat} \& \emph{farm}) & \textbf{or} \\ & (\emph{wheat} \&
\emph{commodity}) & \textbf{or} \\ & (\emph{bushels} \& \emph{export})
& \textbf{or} \\ & (\emph{wheat} \& \emph{tonnes}) & \textbf{or} \\ &
(\emph{wheat} \& \emph{winter} \& $\neg$ \emph{soft})) & \textbf{then}
\ \textsc{Wheat} \ \textbf{else} \ $\neg$ \textsc{Wheat} \\ \hline
\end{tabular}\caption{\label{fig:construe}Rule-based classifier for
the \textsc{Wheat} category; keywords are indicated in \emph{italic},
categories are indicated in \textsc{Small Caps} (from
\protect\cite{Apte94}).}\end{center}\end{figure}

The drawback of this approach is the \emph{knowledge acquisition
bottleneck} well-known from the expert systems literature. That is,
the rules must be manually defined by a knowledge engineer with the
aid of a domain expert (in this case, an expert in the membership of
documents in the chosen set of categories): if the set of categories
is updated, then these two professionals must intervene again, and if
the classifier is ported to a completely different domain (i.e.\ set
of categories) a different domain expert needs to intervene and the
work has to be repeated from scratch.

On the other hand, it was originally suggested that this approach can
give very good effectiveness results: Hayes et al.\
\citeyear{Hayes90a} reported a .90 ``breakeven'' result (see Section
\ref{sec:evaluation}) on a subset of the {\sf Reuters} test
collection, a figure that outperforms even the best classifiers built
in the late '90s by state-of-the-art ML techniques. However, no other
classifier has been tested on the same dataset as \textsc{Construe},
and it is not clear whether this was a randomly chosen or a favourable
subset of the entire {\sf Reuters} collection. As argued in
\cite{Yang99a}, the results above do not allow us to state that these
effectiveness results may be obtained in general.

Since the early '90s, the ML approach to TC has gained popularity and
has eventually become the dominant one, at least in the research
community (see \cite{Mitchell96} for a comprehensive introduction to
ML). In this approach a general inductive process (also called the
\emph{learner}) automatically builds a classifier for a category $c_i$
by observing the characteristics of a set of documents manually
classified under $c_i$ or $\overline{c}_i$ by a domain expert; from
these characteristics, the inductive process gleans the
characteristics that a new unseen document should have in order to be
classified under $c_i$. In ML terminology, the classification problem
is an activity of \emph{supervised} learning, since the learning
process is ``supervised'' by the knowledge of the categories and of
the training instances that belong to them\footnote{Within the area of
content-based document management tasks, an example of an {\em
unsupervised} learning activity is \emph{document clustering} (see
Section \ref{sec:intro}).}.

The advantages of the ML approach over the KE approach are evident.
The engineering effort goes towards the construction not of a
classifier, but of an automatic builder of classifiers (the
\emph{learner}). This means that if a learner is (as it often is)
available off-the-shelf, all that is needed is the inductive,
\emph{automatic} construction of a classifier from a set of manually
classified documents. The same happens if a classifier already exists
and the original set of categories is updated, or if the classifier is
ported to a completely different domain.

In the ML approach the preclassified documents are then the key
resource. In the most favourable case they are already available; this
typicaly happens for organizations which have previously carried out
the same categorization activity manually and decide to automate the
process. The less favourable case is when no manually classified
documents are available; this typicaly happens for organizations which
start a categorization activity and opt for an automated modality
straightaway. The ML approach is more convenient than the KE approach
also in this latter case. In fact, it is easier to manually classify a
set of documents than to build and tune a set of rules, since it is
easier to characterize a concept extensionally (i.e.\ to select
instances of it) than intensionally (i.e.\ to describe the concept in
words, or to describe a procedure for recognizing its instances).

Classifiers built by means of ML techniques nowadays achieve
impressive levels of effectiveness (see Section \ref{sec:evaluation}),
making automatic classification a \emph{qualitatively} (and not only
economically) viable alternative to manual classification.


\subsection{Training set, test set, and validation set}
\label{sec:trainingandtest}
 
The ML approach relies on the availability of an \emph{initial corpus}
$\Omega =\{d_1, \ldots, d_{|\Omega |}\} \subset {\cal D}$ of documents
preclassified under ${\cal C}=\{c_1, \ldots, c_{|{\cal C}|}\}$. That
is, the values of the total function $\breve{\Phi}: {\cal D}\times
{\cal C}\rightarrow\{T,F\}$ are known for every pair $\langle d_j,c_i
\rangle \in \Omega \times {\cal C}$. A document $d_j$ is a {\em
positive example} of $c_i$ if $\breve{\Phi}(d_j,c_i)=T$, a {\em
negative example} of $c_i$ if $\breve{\Phi}(d_j,c_i)=F$.

In research settings (and in most operational settings too), once a
classifier $\Phi$ has been built it is desirable to evaluate its
effectiveness. In this case, prior to classifier construction the
initial corpus is split in two sets, not necessarily of equal size:

\begin{itemize}

\item a \emph{training(-and-validation) set} $TV=\{d_1, \ldots,
d_{|TV|}\}$. The classifier $\Phi$ for categories ${\cal C}=\{c_1,
\ldots, c_{|{\cal C}|}\}$ is inductively built by observing the
characteristics of these documents;

\item a \emph{test set} $Te=\{d_{|TV| +1}, \ldots, d_{| \Omega |}\}$,
used for testing the effectiveness of the classifiers. Each $d_j\in
Te$ is fed to the classifier, and the classifier decisions
$\Phi(d_j,c_i)$ are compared with the expert decisions
$\breve{\Phi}(d_j,c_i)$. A measure of classification effectiveness is
based on how often the $\Phi(d_j,c_i)$ values match the
$\breve{\Phi}(d_j,c_i)$ values.

\end{itemize}

\noindent The documents in $Te$ cannot participate in any way in the
inductive construction of the classifiers; if this condition were not
satisfied the experimental results obtained would likely be
unrealistically good, and the evaluation would thus have no scientific
character \cite[page 129]{Mitchell96}. In an operational setting,
after evaluation has been performed one would typically re-train the
classifier on the entire initial corpus, in order to boost
effectiveness. In this case the results of the previous evaluation
would be a pessimistic estimate of the real performance, since the
final classifier has been trained on more data than the classifier
evaluated.

This is called the \emph{train-and-test} approach. An alternative is
the \emph{$k$-fold cross-validation} approach (see e.g.\ \cite[page
146]{Mitchell96}), in which $k$ different classifiers $\Phi_1, \ldots,
\Phi_k$ are built by partitioning the initial corpus into $k$ disjoint
sets $Te_1, \ldots, Te_k$ and then iteratively applying the
train-and-test approach on pairs $\langle TV_i=\Omega -Te_i,
Te_i\rangle$. The final effectiveness figure is obtained by
individually computing the effectiveness of $\Phi_1, \ldots, \Phi_k$,
and then averaging the individual results in some way.

In both approaches it is often the case that the internal parameters
of the classifiers must be tuned, by testing which values of the
parameters yield the best effectiveness. In order to make this
optimization possible, in the train-and-test approach the set $\{d_1,
\ldots, d_{|TV|}\}$ is further split into a \emph{training set}
$Tr=\{d_1, \ldots, d_{|Tr|}\}$, from which the classifier is built,
and a {\em validation set} $Va=\{d_{|Tr| +1}, \ldots, d_{|TV|}\}$
(sometimes called a \emph{hold-out set}), on which the repeated tests
of the classifier aimed at parameter optimization are performed; the
obvious variant may be used in the $k$-fold cross-validation case.
Note that, for the same reason why we do not test a classifier on the
documents it has been trained on, we do not test it on the documents
it has been optimized on: test set and validation set must be kept
separate\footnote{From now on, we will take the freedom to use the
expression ``test document'' to denote any document not in the
training set and validation set. This includes thus any document
submitted to the classifier in the operational phase.}.

Given a corpus $\Omega$, one may define the \emph{generality}
$g_{\Omega}(c_i)$ of a category $c_i$ as the percentage of documents
that belong to $c_i$, i.e.: \begin{eqnarray}\nonumber g_{\Omega}(c_i)
& = & \frac{|\{d_j \in \Omega \ | \
\breve{\Phi}(d_j,c_i)=T\}|}{|\Omega|}\end{eqnarray}

\noindent The \emph{training set generality} $g_{Tr}(c_i)$, {\em
validation set generality} $g_{Va}(c_i)$, and \emph{test set
generality} $g_{Te}(c_i)$ of $c_i$ may be defined in the obvious way.


\subsection{Information retrieval techniques and text categorization}
\label{sec:IRinclassification}

Text categorization heavily relies on the basic machinery of IR. The
reason is that TC is a content-based document management task, and as
such it shares many characteristics with other IR tasks such as text
search.

IR techniques are used in three phases of the text classifier life
cycle:

\begin{enumerate}

\item \label{item:indexing} IR-style \emph{indexing} is always
performed on the documents of the initial corpus and on those to be
classified during the operational phase;

\item \label{item:construction} IR-style techniques (such as
document-request matching, query reformulation, \ldots) are often used
in the \emph{inductive construction} of the classifiers;

\item \label{item:evaluation} IR-style \emph{evaluation} of the
effectiveness of the classifiers is performed.

\end{enumerate}

\noindent The various approaches to classification differ mostly for
how they tackle (\ref{item:construction}), although in a few cases
non-standard approaches to (\ref{item:indexing}) and
(\ref{item:evaluation}) are also used. Indexing, induction and
evaluation are the themes of Sections \ref{sec:indexingandDR},
\ref{sec:construction} and \ref{sec:evaluation}, respectively.


\section{Document indexing and dimensionality reduction}
\label{sec:indexingandDR}

\subsection{Document indexing}\label{sec:indexing}

Texts cannot be directly interpreted by a classifier or by a
classifier-building algorithm. Because of this, an \emph{indexing}
procedure that maps a text $d_j$ into a compact representation of its
content needs to be uniformly applied to training, validation and test
documents. The choice of a representation for text depends on what one
regards as the meaningful units of text (the problem of \emph{lexical
semantics}) and the meaningful natural language rules for the
combination of these units (the problem of \emph{compositional
semantics}). Similarly to what happens in IR, in TC this latter
problem is usually disregarded\footnote{An exception to this is
represented by learning approaches based on \emph{Hidden Markov
Models}~\cite{Denoyer01,Frasconi01}.}, and a text $d_j$ is usually
represented as a vector of term \emph{weights} $\vec{d}_j=\langle
w_{1j}, \ldots, w_{|{\cal T}| j}\rangle$, where ${\cal T}$ is the set
of \emph{terms} (sometimes called \emph{features}) that occur at least
once in at least one document of $Tr$, and $0\leq w_{kj}\leq 1$
represents, loosely speaking, how much term $t_k$ contributes to the
semantics of document $d_j$. Differences among approaches are
accounted for by

\begin{enumerate}

\item \label{item:indexterm} different ways to understand what a term
is;

\item \label{item:weight} different ways to compute term weights.

\end{enumerate}

\noindent A typical choice for (\ref{item:indexterm}) is to identify
terms with words. This is often called either the \emph{set of words}
or the \emph{bag of words} approach to document representation,
depending on whether weights are binary or not.

In a number of experiments \cite{Apte94,Dumais98,Lewis92} it has been
found that representations more sophisticated than this do not yield
significantly better effectiveness, thereby confirming similar results
from IR \cite{Salton88}. In particular, some authors have used
\emph{phrases}, rather than individual words, as indexing terms
\cite{Fuhr91a,Schutze95,Tzeras93}, but the experimental results found
to date have not been uniformly encouraging, irrespectively of whether
the notion of ``phrase'' is motivated

\begin{itemize}

\item \emph{syntactically}, i.e.\ the phrase is such according to a
grammar of the language (see e.g.\ \cite{Lewis92});

\item \emph{statistically}, i.e.\ the phrase is not grammatically
such, but is composed of a set/sequence of words whose patterns of
contiguous occurrence in the collection are statistically significant
(see e.g.\ \cite{Caropreso01}).

\end{itemize}

\noindent Lewis \citeyear{Lewis92} argues that the likely reason for
the discouraging results is that, although indexing languages based on
phrases have superior semantic qualities, they have inferior
statistical qualities with respect to word-only indexing languages: a
phrase-only indexing language has ``more terms, more synonymous or
nearly synonymous terms, lower consistency of assignment (since
synonymous terms are not assigned to the same documents), and lower
document frequency for terms'' \cite[page 40]{Lewis92}. Although his
remarks are about syntactically motivated phrases, they also apply to
statistically motivated ones, although perhaps to a smaller degree. A
combination of the two approaches is probably the best way to go:
Tzeras and Hartmann \citeyear{Tzeras93} obtained significant
improvements by using noun phrases obtained through a combination of
syntactic and statistical criteria, where a ``crude'' syntactic method
was complemented by a statistical filter (only those syntactic phrases
that occurred at least three times in the positive examples of a
category $c_i$ were retained). It is likely that the final word on the
usefulness of phrase indexing in TC has still to be told, and
investigations in this direction are still being actively pursued
\cite{Caropreso01,Mladenic98d}.

As for issue (\ref{item:weight}), weights usually range between 0 and
1 (an exception is \cite{Lewis96}), and for ease of exposition we will
assume they always do. As a special case, binary weights may be used
(1 denoting presence and 0 absence of the term in the document);
whether binary or non-binary weights are used depends on the
classifier learning algorithm used. In the case of non-binary
indexing, for determining the weight $w_{kj}$ of term $t_k$ in
document $d_j$ any IR-style indexing technique that represents a
document as a vector of weighted terms may be used. Most of the times,
the standard $tfidf$ function is used (see e.g.\ \cite{Salton88}),
defined as \begin{eqnarray} \label{eqn:tfidf} tfidf(t_k,d_j) & =
\#(t_k,d_j)\cdot \log \displaystyle\frac{|Tr|}{\#_{Tr}(t_k)}
\end{eqnarray}

\noindent where $\#(t_k,d_j)$ denotes the number of times $t_k$ occurs
in $d_j$, and $\#_{Tr}(t_k)$ denotes the \emph{document frequency} of
term $t_k$, i.e.\ the number of documents in $Tr$ in which $t_k$
occurs. This function embodies the intuitions that (i) the more often
a term occurs in a document, the more it is representative of its
content, and (ii) the more documents a term occurs in, the less
discriminating it is\footnote{There exist many variants of $tfidf$,
that differ from each other in terms of logarithms, normalization or
other correction factors. Formula \ref{eqn:tfidf} is just one of the
possible instances of this class; see \cite{Salton88,Singhal96} for
variations on this theme.}. Note that this formula (as most other
indexing formulae) weights the importance of a term to a document in
terms of occurrence considerations only, thereby deeming of null
importance the order in which the terms occur in the document and the
syntactic role they play. In other words, the semantics of a document
is reduced to the collective lexical semantics of the terms that occur
in it, thereby disregarding the issue of compositional semantics (an
exception are the representation techniques used for {\sc Foil}
\cite{Cohen95a} and \textsc{Sleeping Experts} \cite{Cohen99}).

In order for the weights to fall in the [0,1] interval and for the
documents to be represented by vectors of equal length, the weights
resulting from $tfidf$ are often normalized by \emph{cosine
normalization}, given by:
\begin{eqnarray}\label{eq:cosinenormalization}w_{kj}& = &
\displaystyle\frac{tfidf(t_k,d_j)}{\sqrt{\sum_{s=1}^{|{\cal
T}|}(tfidf(t_s,d_j))^2}}\end{eqnarray}

\noindent Although normalized $tfidf$ is the most popular one, other
indexing functions have also been used, including probabilistic
techniques \cite{Goevert99} or techniques for indexing structured
documents \cite{Larkey96}. Functions different from $tfidf$ are
especially needed when $Tr$ is not available in its entirety from the
start and $\#_{Tr}(t_k)$ cannot thus be computed, as e.g.\ in adaptive
filtering; in this case approximations of $tfidf$ are usually employed
\cite[Section 4.3]{Dagan97}.

Before indexing, the removal of \emph{function words} (i.e.\
topic-neutral words such as articles, prepositions, conjunctions,
etc.) is almost always performed (exceptions include
\cite{Lewis96,Nigam00,Riloff95})\footnote{One application of TC in
which it would be inappropriate to remove function words is author
identification for documents of disputed paternity. In fact, as noted
in \cite[page 589]{Manning99}, ``it is often the `little' words that
give an author away (for example, the relative frequencies of words
like {\sf because} or {\sf though})''.}. Concerning {\em stemming}
(i.e.\ grouping words that share the same morphological root), its
suitability to TC is controversial. Although, similarly to
unsupervised term clustering (see Section \ref{sec:termclustering}) of
which it is an instance, stemming has sometimes been reported to hurt
effectiveness (e.g.\ \cite{Baker98}), the recent tendency is to adopt
it, as it reduces both the dimensionality of the term space (see
Section \ref{sec:dimensionalityreduction}) and the stochastic
dependence between terms (see Section \ref{sec:probabilistic}).
 
Depending on the application, either the full text of the document or
selected parts of it are indexed. While the former option is the rule,
exceptions exist. For instance, in a patent categorization application
Larkey \citeyear{Larkey99} indexes only the title, the abstract, the
first 20 lines of the summary, and the section containing the claims
of novelty of the described invention. This approach is made possible
by the fact that documents describing patents are structured.
Similarly, when a document title is available, one can pay extra
importance to the words it contains \cite{Apte94,Cohen99,Weiss99}.
When documents are flat, identifying the most relevant part of a
document is instead a non-obvious task.


\subsection{The Darmstadt Indexing Approach}
\label{sec:DIA}

The AIR/X system \cite{Fuhr91a} occupies a special place in the
literature on indexing for TC. This system is the final result of the
AIR project, one of the most important efforts in the history of TC:
spanning a duration of more than ten years \cite{Knorz82,Tzeras93}, it
has produced a system operatively employed since 1985 in the
classification of corpora of scientific literature of $O(10^5)$
documents and $O(10^4)$ categories, and has had important theoretical
spin-offs in the field of probabilistic indexing
\cite{Fuhr89,Fuhr91}\footnote{The AIR/X system, its applications
(including the AIR/PHYS system \cite{Biebricher88}, an application of
AIR/X to indexing physics literature), and its experiments, have also
been richly documented in a series of papers and doctoral theses
written in German. The interested reader may consult \cite{Fuhr91a}
for a detailed bibliography.}.

The approach to indexing taken in AIR/X is known as the
\emph{Darmstadt Indexing Approach} (DIA) \cite{Fuhr85}. Here,
``indexing'' is used in the sense of Section
\ref{sec:indexingforBoolean}, i.e.\ as using terms from a controlled
vocabulary, and is thus a synonym of TC (the DIA was later extended to
indexing with free terms \cite{Fuhr91}). The idea that underlies the
DIA is the use of a much wider set of ``features'' than described in
Section \ref{sec:indexing}. All other approaches mentioned in this
paper view \emph{terms} as the dimensions of the learning space, where
terms may be single words, stems, phrases, or (see Sections
\ref{sec:termclustering} and \ref{sec:latentsemanticindexing})
combinations of any of these. In contrast, the DIA considers
\emph{properties} (of terms, documents, categories, or pairwise
relationships among these) as basic dimensions of the learning space.
Examples of these are

\begin{itemize}

\item \emph{properties of a term $t_k$}: e.g.\ the $idf$ of $t_k$;

\item \emph{properties of the relationship between a term $t_k$ and a
document $d_j$}: e.g.\ the $tf$ of $t_k$ in $d_j$; or the location
(e.g.\ in the title, or in the abstract) of $t_k$ within $d_j$;

\item \emph{properties of a document $d_j$}: e.g.\ the length of
$d_j$;

\item \emph{properties of a category $c_i$}: e.g.\ the training set
generality of $c_i$.

\end{itemize}

\noindent For each possible document-category pair, the values of
these features are collected in a so-called \emph{relevance
description} vector $\vec{rd}(d_j, c_i)$. The size of this vector is
determined by the number of properties considered, and is thus
independent of specific terms, categories or documents (for
multivalued features, appropriate aggregation functions are applied in
order to yield a single value to be included in $\vec{rd}(d_j, c_i)$);
in this way an abstraction from specific terms, categories or
documents is achieved.

The main advantage of this approach is the possibility to consider
additional features that can hardly be accounted for in the usual
term-based approaches, e.g.\ the location of a term within a document,
or the certainty with which a phrase was identified in a document. The
term-category relationship is described by estimates, derived from the
training set, of the probability $P(c_i|t_k)$ that a document belongs
to category $c_i$, given that it contains term $t_k$ (the \emph{DIA
association factor})\footnote{Association factors are called {\em
adhesion coefficients} in many early papers on TC; see e.g.\
\cite{Field75,Robertson84}.}. Relevance description vectors
$\vec{rd}(d_j, c_i)$ are then the final representations that are used
for the classification of document $d_j$ under category $c_i$.

The essential ideas of the DIA -- transforming the classification
space by means of abstraction and using a more detailed text
representation than the standard bag-of-words approach -- have not
been taken up by other researchers so far. For new TC applications
dealing with structured documents or categorization of Web pages,
these ideas may become of increasing importance.


\subsection{Dimensionality reduction}
\label{sec:dimensionalityreduction}

Unlike in text retrieval, in TC the high dimensionality of the term
space (i.e.\ the large value of $|{\cal T}|$) may be problematic. In
fact, while typical algorithms used in text retrieval (such as cosine
matching) can scale to high values of $|{\cal T}|$, the same does not
hold of many sophisticated learning algorithms used for classifier
induction (e.g.\ the LLSF algorithm of \cite{Yang94}). Because of
this, before classifier induction one often applies a pass of {\em
dimensionality reduction} (DR), whose effect is to reduce the size of
the vector space from $|{\cal T}|$ to $|{\cal T}'|\ll |{\cal T}|$; the
set ${\cal T}'$ is called the \emph{reduced term set}.

DR is also beneficial since it tends to reduce \emph{overfitting},
i.e.\ the phenomenon by which a classifier is tuned also to the {\em
contingent} characteristics of the training data rather than just the
\emph{constitutive} characteristics of the categories. Classifiers
which overfit the training data are good at re-classifying the data
they have been trained on, but much worse at classifying previously
unseen data. Experiments have shown that in order to avoid overfitting
a number of training examples roughly proportional to the number of
terms used is needed; Fuhr and Buckley \citeyear[page 235]{Fuhr91}
have suggested that 50-100 training examples per term may be needed in
TC tasks. This means that if DR is performed, overfitting may be
avoided even if a smaller amount of training examples is used.
However, in removing terms the risk is to remove potentially useful
information on the meaning of the documents. It is then clear that, in
order to obtain optimal (cost-)effectiveness, the reduction process
must be performed with care. Various DR methods have been proposed,
either from the information theory or from the linear algebra
literature, and their relative merits have been tested by
experimentally evaluating the variation in effectiveness that a given
classifier undergoes after application of the function to the term
space.

There are two distinct ways of viewing DR, depending on whether the
task is performed locally (i.e.\ for each individual category) or
globally:

\begin{itemize}

\item \label{item:locally} \emph{local DR}: for each category $c_i$, a
set ${\cal T}_{i}'$ of terms, with $|{\cal T}_{i}'|\ll |{\cal T}|$, is
chosen for classification under $c_i$ (see e.g.\
\cite{Apte94,Lewis94,Li98a,Ng97,Sable00,Schutze95,Wiener95}). This
means that different subsets of $\vec{d}_j$ are used when working with
the different categories. Typical values are $10\leq |{\cal
T}_{i}'|\leq 50$.

\item \label{item:globally} \emph{global DR}: a set ${\cal T}'$ of
terms, with $|{\cal T}'|\ll|{\cal T}|$, is chosen for the
classification under all categories ${\cal C}=\{c_1, \ldots, c_{|{\cal
C}|}\}$ (see e.g.\ \cite{Caropreso01,Mladenic98b,Yang99a,Yang97}).

\end{itemize}

\noindent This distinction usually does not impact on the choice of DR
technique, since most such techniques can be used (and have been used)
for local and global DR alike (\emph{supervised} DR techniques -- see
Section \ref{sec:termclustering} -- are exceptions to this rule). In
the rest of this section we will assume that the global approach is
used, although everything we will say also applies to the local
approach.

A second, orthogonal distinction may be drawn in terms of the nature
of the resulting terms:

\begin{itemize}

\item \label{item:termselection} \emph{DR by term selection}: ${\cal
T}'$ is a subset of ${\cal T}$;

\item \label{item:termextraction} \emph{DR by term extraction}: the
terms in ${\cal T}'$ are not of the same type of the terms in ${\cal
T}$ (e.g.\ if the terms in ${\cal T}$ are words, the terms in ${\cal
T}'$ may not be words at all), but are obtained by combinations or
transformations of the original ones.

\end{itemize}

\noindent Unlike in the previous distinction, these two ways of doing
DR are tackled by very different techniques; we will address them
separately in the next two sections.


\subsection{Dimensionality reduction by term selection}
\label{sec:termselection}

Given a predetermined integer $r$, techniques for term selection (also
called \emph{term space reduction} -- TSR) attempt to select, from the
original set ${\cal T}$, the set ${\cal T}'$ of terms (with $|{\cal
T}'|\ll|{\cal T}|$) that, when used for document indexing, yields the
highest effectiveness. Yang and Pedersen \citeyear{Yang97} have shown
that TSR may even result in a moderate ($\leq 5\%$) increase in
effectiveness, depending on the classifier, on the \emph{aggressivity}
$\frac{|{\cal T}|}{|{\cal T}'|}$ of the reduction, and on the TSR
technique used.

Moulinier et al.\ \citeyear{Moulinier96} have used a so-called {\em
wrapper} approach, i.e.\ one in which ${\cal T}'$ is identified by
means of the same learning method which will be used for building the
classifier \cite{John94}. Starting from an initial term set, a new
term set is generated by either adding or removing a term. When a new
term set is generated, a classifier based on it is built and then
tested on a validation set. The term set that results in the best
effectiveness is chosen. This approach has the advantage of being
tuned to the learning algorithm being used; moreover, if local DR is
performed, different numbers of terms for different categories may be
chosen, depending on whether a category is or is not easily separable
from the others. However, the sheer size of the space of different
term sets makes its cost prohibitive for standard TC applications.

A computationally easier alternative is the \emph{filtering} approach
\cite{John94}, i.e.\ keeping the $|{\cal T}'|\ll |{\cal T}|$ terms
that receive the highest score according to a function that measures
the ``importance'' of the term for the TC task. We will explore this
solution in the rest of this section.


\subsubsection{Document frequency}
\label{sec:documentfrequency}

A simple and effective global TSR function is the \emph{document
frequency} $\#_{Tr}(t_k)$ of a term $t_k$, i.e.\ only the terms that
occur in the highest number of documents are retained. In a series of
experiments Yang and Pedersen \citeyear{Yang97} have shown that with
$\#_{Tr}(t_k)$ it is possible to reduce the dimensionality by a factor
of 10 with no loss in effectiveness (a reduction by a factor of 100
bringing about just a small loss).

This seems to indicate that the terms occurring most frequently in the
collection are the most valuable for TC. As such, this would seem to
contradict a well-known ``law'' of IR, according to which the terms
with low-to-medium document frequency are the most informative ones
\cite{Salton88}. But these two results do not contradict each other,
since it is well-known (see e.g.\ \cite{Salton75}) that the large
majority of the words occurring in a corpus have a {\em very} low
document frequency; this means that by reducing the term set by a
factor of 10 using document frequency, only such words are removed,
while the words from low-to-medium to high document frequency are
preserved. Of course, stop words need to be removed in advance, lest
only topic-neutral words are retained \cite{Mladenic98b}.

Finally, note that a slightly more empirical form of TSR by document
frequency is adopted by many authors, who remove all terms occurring
in at most $x$ training documents (popular values for $x$ range from 1
to 3), either as the only form of DR \cite{Maron61,Ittner95} or before
applying another more sophisticated form \cite{Dumais98,Li98a}. A
variant of this policy is removing all terms that occur at most $x$
times in the training set (e.g.\ \cite{Dagan97,Joachims97}), with
popular values for $x$ ranging from 1 (e.g.\ \cite{Baker98}) to 5
(e.g.\ \cite{Apte94,Cohen95a}).


\subsubsection{Other information-theoretic term selection functions}
\label{sec:informationtheoretic}

Other more sophisticated information-theoretic functions have been
used in the literature, among which the \emph{DIA association factor}
\cite{Fuhr91a}, {\em chi-square}
\cite{Caropreso01,Galavotti00,Schutze95,Sebastiani00,Yang97,Yang99},
\emph{NGL coefficient} \cite{Ng97,Ruiz99}, \emph{information gain}
\cite{Caropreso01,Larkey98,Lewis92,Lewis94,Mladenic98b,Moulinier96a,Yang97,Yang99},
\emph{mutual information}
\cite{Dumais98,Lam97,Larkey96,Lewis94,Li98a,Moulinier96,Ruiz99,Taira99,Yang97},
\emph{odds ratio} \cite{Caropreso01,Mladenic98b,Ruiz99},
\emph{relevancy score} \cite{Wiener95}, and \emph{GSS coefficient}
\cite{Galavotti00}. The mathematical definitions of these measures are
summarized for convenience in Table
\ref{tab:termspacereduction}\footnote{For better uniformity Table
\ref{tab:termspacereduction} views all the TSR functions in terms of
subjective probability. In some cases such as $\#(t_k,c_i)$ and
$\chi^2(t_k,c_i)$ this is slightly artificial, since these two
functions are not usually viewed in probabilistic terms. The formulae
refer to the ``local'' (i.e.\ category-specific) forms of the
functions, which again is slightly artificial in some cases (e.g.\
$\#(t_k,c_i)$). Note that the NGL and GSS coefficients are here named
after their authors, since they had originally been given names that
might generate some confusion if used here.}. Here, probabilities are
interpreted on an event space of documents (e.g.\
$P(\overline{t}_k,c_i)$ denotes the probability that, for a random
document $x$, term $t_k$ does not occur in $x$ and $x$ belongs to
category $c_i$), and are estimated by counting occurrences in the
training set. All functions are specified ``locally'' to a specific
category $c_i$; in order to assess the value of a term $t_k$ in a
``global'', category-independent sense, either the sum
$f_{sum}(t_k)=\sum_{i=1}^{|{\cal C}|}f(t_k,c_i)$, or the weighted
average $f_{wavg}(t_k)=\sum_{i=1}^{|{\cal C}|}P(c_i)f(t_k,c_i)$, or
the maximum $f_{max}(t_k)=\max_{i=1}^{|{\cal C}|}f(t_k,c_i)$ of their
category-specific values $f(t_k,c_i)$ are usually computed.

These functions try to capture the intuition that the best terms for
$c_i$ are the ones distributed most differently in the sets of
positive and negative examples of $c_i$. However, interpretations of
this principle vary across different functions. For instance, in the
experimental sciences $\chi^2$ is used to measure how the results of
an observation differ (i.e.\ are independent) from the results
expected according to an initial hypothesis (lower values indicate
lower dependence). In DR we measure how independent $t_k$ and $c_i$
are. The terms $t_k$ with the lowest value for $\chi^2(t_k,c_i)$ are
thus the most independent from $c_i$; since we are interested in the
terms which are not, we select the terms for which $\chi^2(t_k,c_i)$
is highest.

\begin{table} \begin{center}\begin{tabular}{|c|c|c|} \hline
Function & Denoted by & Mathematical form \\ \hline\hline
\rule[-3ex]{0mm}{7ex} \emph{Document frequency} & $\#(t_k,c_i)$ &
$P(t_k|c_i)$ \\
\hline \rule[-3ex]{0mm}{7ex} \emph{DIA association factor} &
$z(t_k,c_i)$ & $P(c_i|t_k)$ \\ \hline
\rule[-3ex]{0mm}{7ex} \emph{Information gain} & $IG(t_k,c_i)$ &
$\displaystyle\sum_{c\in\{c_i,\overline{c}_i\}}
\displaystyle\sum_{t\in\{t_k,\overline{t}_k\}} P(t,c) \cdot\log
\displaystyle\frac{P(t,c)}{P(t)\cdot P(c)}$ \\ \hline
\rule[-3ex]{0mm}{7ex} \emph{Mutual information} & $MI(t_k,c_i)$ &
$\log \displaystyle\frac{P(t_k,c_i)}{P(t_k)\cdot P(c_i)}$ \\ \hline
\rule[-3ex]{0mm}{7ex} \emph{Chi-square} & $\chi^2(t_k,c_i)$ &
$\displaystyle\frac{|Tr| \cdot [P(t_k,c_i)\cdot
P(\overline{t}_k,\overline{c}_i) - P(t_k,\overline{c}_i)\cdot
P(\overline{t}_k,c_i)]^2}{P(t_k)\cdot P(\overline{t}_k)\cdot
P(c_i)\cdot P(\overline{c}_i)}$ \\ \hline
\rule[-3ex]{0mm}{7ex} \emph{NGL coefficient} & $NGL(t_k,c_i)$ &
$\displaystyle\frac{\sqrt{|Tr|} \cdot [P(t_k,c_i)\cdot
P(\overline{t}_k,\overline{c}_i) - P(t_k,\overline{c}_i)\cdot
P(\overline{t}_k,c_i)]}{\sqrt{P(t_k)\cdot P(\overline{t}_k)\cdot
P(c_i)\cdot P(\overline{c}_i)}}$ \\
\hline \rule[-3ex]{0mm}{7ex} \emph{Relevancy score} & $RS(t_k,c_i)$ &
$\log \displaystyle\frac{P(t_k|c_i)
+d}{P(\overline{t}_k|\overline{c}_i)+d}$ \\ \hline
\rule[-3ex]{0mm}{7ex} \emph{Odds Ratio} & $OR(t_k,c_i)$ &
$\displaystyle\frac{P(t_k|c_i)\cdot
(1-P(t_k|\overline{c}_i))}{(1-P(t_k|c_i))\cdot P(t_k|\overline{c}_i)}$
\\ \hline \rule[-3ex]{0mm}{7ex} \emph{GSS coefficient} &
$GSS(t_k,c_i)$ & $P(t_k, c_i)\cdot P(\overline{t}_k, \overline{c}_i) -
P(t_k, \overline{c}_i)\cdot P(\overline{t}_k, c_i)$ \\ \hline
\end{tabular}\end{center}\caption{\label{tab:termspacereduction}Main
functions used for term space reduction purposes. Information gain is
also known as \emph{expected mutual information}; it is used under
this name by Lewis \protect\citeyear[page 44]{Lewis92} and Larkey
\protect\citeyear{Larkey98}. In the $RS(t_k,c_i)$ formula $d$ is a
constant damping factor.}
\end{table}

While each TSR function has its own rationale, the ultimate word on
its value is the effectiveness it brings about. Various experimental
comparisons of TSR functions have thus been carried out
\cite{Caropreso01,Galavotti00,Mladenic98b,Yang97}. In these
experiments most functions listed in Table
\ref{tab:termspacereduction} (with the possible exception of $MI$)
have improved on the results of document frequency. For instance, Yang
and Pedersen \citeyear{Yang97} have shown that, with various
classifiers and various initial corpora, sophisticated techniques such
as $IG_{sum}(t_k,c_i)$ or $\chi^2_{max}(t_k,c_i)$ can reduce the
dimensionality of the term space by a factor of 100 with no loss (or
even with a small increase) of effectiveness. Collectively, the
experiments reported in the above-mentioned papers seem to indicate
that \{$OR_{sum}$, $NGL_{sum}$, $GSS_{max}$\} $>$ \{$\chi^2_{max}$,
$IG_{sum}$\} $>$ \{$\#_{wavg}$, $\chi^2_{wavg}$\} $\gg$ \{$MI_{max}$,
$MI_{wavg}$\}, where ``$>$'' means ``performs better than''. However,
it should be noted that these results are just indicative, and that
more general statements on the relative merits of these functions
could be made only as a result of comparative experiments performed in
thoroughly controlled conditions and on a variety of different
situations (e.g.\ different classifiers, different initial corpora,
\ldots).


\subsection{Dimensionality reduction by term extraction}
\label{sec:termextraction}

Given a predetermined $|{\cal T}'|\ll |{\cal T}|$, \emph{term
extraction} attempts to generate, from the original set ${\cal T}$, a
set ${\cal T}'$ of ``synthetic'' terms that maximize effectiveness.
The rationale for using synthetic (rather than naturally occurring)
terms is that, due to the pervasive problems of polysemy, homonymy and
synonymy, the original terms may not be optimal dimensions for
document content representation. Methods for term extraction try to
solve these problems by creating artificial terms that do not suffer
from them. Any term extraction method consists in (i) a method for
extracting the new terms from the old ones, and (ii) a method for
converting the original document representations into new
representations based on the newly synthesized dimensions. Two term
extraction methods have been experimented in TC, namely term
clustering and latent semantic indexing.


\subsubsection{Term clustering} \label{sec:termclustering}

\emph{Term clustering} tries to group words with a high degree of
pairwise semantic relatedness, so that the groups (or their centroids,
or a representative of them) may be used instead of the terms as
dimensions of the vector space. Term clustering is different from term
selection, since the former tends to address terms \emph{synonymous}
(or near-synonymous) with other terms, while the latter targets
\emph{non-informative} terms\footnote{Some term selection methods,
such as wrapper methods, also address the problem of redundancy.}.

Lewis \citeyear{Lewis92} was the first to investigate the use of term
clustering in TC. The method he employed, called \emph{reciprocal
nearest neighbour clustering}, consists in creating clusters of two
terms that are one the most similar to the other according to some
measure of similarity. His results were inferior to those obtained by
single-word indexing, possibly due to a disappointing performance by
the clustering method: as Lewis \citeyear[page 48]{Lewis92} says,
``The relationships captured in the clusters are mostly accidental,
rather than the systematic relationships that were hoped for.''

Li and Jain \citeyear{Li98a} view semantic relatedness between words
in terms of their co-occurrence and co-absence within training
documents. By using this technique in the context of a hierarchical
clustering algorithm they witnessed only a marginal effectiveness
improvement; however, the small size of their experiment (see Section
\ref{sec:committees}) hardly allows any definitive conclusion to be
reached.

Both \cite{Lewis92,Li98a} are examples of \emph{unsupervised}
clustering, since clustering is not affected by the category labels
attached to the documents. Baker and McCallum \citeyear{Baker98}
provide instead an example of \emph{supervised} clustering, as the
\emph{distributional clustering} method they employ clusters together
those terms that tend to indicate the presence of the same category,
or group of categories. Their experiments, carried out in the context
of a Na\"{\i}ve Bayes classifier (see Section
\ref{sec:probabilistic}), showed only a 2\% effectiveness loss with an
aggressivity of 1000, and even showed some effectiveness improvement
with less aggressive levels of reduction. Later experiments by Slonim
and Tishby \citeyear{Slonim01} have confirmed the potential of
supervised clustering methods for term extraction.


\subsubsection{Latent semantic indexing}
\label{sec:latentsemanticindexing}

\emph{Latent semantic indexing} (LSI -- \cite{Deerwester90}) is a DR
technique developed in IR in order to address the problems deriving
from the use of synonymous, near-synonymous and polysemous words as
dimensions of document representations. This technique compresses
document vectors into vectors of a lower-dimensional space whose
dimensions are obtained as combinations of the original dimensions by
looking at their patterns of co-occurrence. In practice, LSI infers
the dependence among the original terms from a corpus and ``wires''
this dependence into the newly obtained, independent dimensions. The
function mapping original vectors into new vectors is obtained by
applying a singular value decomposition to the matrix formed by the
original document vectors. In TC this technique is applied by deriving
the mapping function from the training set and then applying it to
training and test documents alike.

One characteristic of LSI is that the newly obtained dimensions are
not, unlike in term selection and term clustering, intuitively
interpretable. However, they work well in bringing out the ``latent''
semantic structure of the vocabulary used in the corpus. For instance,
Sch\"utze et al.\ \citeyear[page 235]{Schutze95} discuss the
classification under category {\tt Demographic shifts in the U.S.\
with economic impact} of a document that was indeed a positive test
instance for the category, and that contained, among others, the quite
revealing sentence ``{\tt The nation grew to 249.6 million people in
the 1980s as more Americans left the industrial and
ag\-ri\-cul\-tur\-al heartlands for the South and West}''. The
classifier decision was incorrect when local DR had been performed by
$\chi^2$-based term selection retaining the top original 200 terms,
but was correct when the same task was tackled by means of LSI. This
well exemplifies how LSI works: the above sentence does not contain
any of the 200 terms most relevant to the category selected by
$\chi^2$, but quite possibly the words contained in it have concurred
to produce one or more of the LSI higher-order terms that generate the
document space of the category. As Sch\"utze et al.\ \citeyear[page
230]{Schutze95} put it, ``if there is a great number of terms which
all contribute a small amount of critical information, then the
combination of evidence is a major problem for a term-based
classifier''. A drawback of LSI, though, is that if some original term
is particularly good in itself at discriminating a category, that
discrimination power may be lost in the new vector space.

Wiener et al.\ \citeyear{Wiener95} use LSI in two alternative ways:
(i) for local DR, thus creating several category-specific LSI
representations, and (ii) for global DR, thus creating a single LSI
representation for the entire category set. Their experiments showed
the former approach to perform better than the latter, and both
approaches to perform better than simple TSR based on Relevancy Score
(see Table \ref{tab:termspacereduction}).

Sch\"utze et al.\ \citeyear{Schutze95} experimentally compared
LSI-based term extraction with $\chi^2$-based TSR using three
different classifier learning techniques (namely, linear discriminant
analysis, logistic regression and neural networks). Their experiments
showed LSI to be far more effective than $\chi^2$ for the first two
techniques, while both methods performed equally well for the neural
network classifier.

For other TC works that use LSI or similar term extraction techniques
see e.g.\ \cite{Hull94,Li98a,Schutze98,Weigend99,Yang95}.


\section{Inductive construction of text classifiers}
\label{sec:construction}

The inductive construction of text classifiers has been tackled in a
variety of ways. Here we will deal only with the methods that have
been most popular in TC, but we will also briefly mention the
existence of alternative, less standard approaches.

We start by discussing the general form that a text classifier has.
Let us recall from Section \ref{sec:catandrank} that there are two
alternative ways of viewing classification: ``hard'' (fully automated)
classification and ranking (semi-automated) classification.

The inductive construction of a ranking classifier for category
$c_i\in {\cal C}$ usually consists in the definition of a function
$CSV_i:{\cal D}\rightarrow [0,1]$ that, given a document $d_j$,
returns a \emph{categorization status value} for it, i.e.\ a number
between 0 and 1 that, roughly speaking, represents the evidence for
the fact that $d_j \in c_i$. Documents are then ranked according to
their $CSV_i$ value. This works for ``document-ranking TC'';
``category-ranking TC'' is usually tackled by ranking, for a given
document $d_j$, its $CSV_i$ scores for the different categories in
${\cal C}=\{c_1, \ldots, c_{|{\cal C}|}\}$.

The $CSV_i$ function takes up different meanings according to the
learning method used: for instance, in the ``Na\"{\i}ve Bayes''
approach of Section \ref{sec:probabilistic} $CSV_i(d_j)$ is defined in
terms of a probability, whereas in the ``Rocchio'' approach discussed
in Section \ref{sec:Rocchio} $CSV_i(d_j)$ is a measure of vector
closeness in $|{\cal T}|$-dimensional space.

The construction of a ``hard'' classifier may follow two alternative
paths. The former consists in the definition of a function
$CSV_i:{\cal D}\rightarrow \{T,F\}$. The latter consists instead in
the definition of a function $CSV_i:{\cal D}\rightarrow [0,1]$,
analogous to the one used for ranking classification, followed by the
definition of a \emph{threshold} $\tau_i$ such that
$CSV_i(d_j)\geq\tau_i$ is interpreted as $T$ while $CSV_i(d_j)<\tau_i$
is interpreted as $F$\footnote{Alternative methods are possible, such
as training a classifier for which some standard, predefined value
such as 0 is the threshold. For ease of exposition we will not discuss
them.}.

The definition of thresholds will be the topic of Section
\ref{sec:thresholding}. In Sections \ref{sec:probabilistic} to
\ref{sec:other} we will instead concentrate on the definition of
$CSV_i$, discussing a number of approaches that have been applied in
the TC literature. In general we will assume we are dealing with
``hard'' classification; it will be evident from the context how and
whether the approaches can be adapted to ranking classification. The
presentation of the algorithms will be mostly qualitative rather than
quantitative, i.e.\ will focus on the methods for classifier learning
rather than on the effectiveness and efficiency of the classifiers
built by means of them; this will instead be the focus of Section
\ref{sec:evaluation}.


\subsection{Determining thresholds} \label{sec:thresholding}

There are various policies for determining the threshold $\tau_i$,
also depending on the constraints imposed by the application. The most
important distinction is whether the threshold is derived {\em
analytically} or \emph{experimentally}.

The former method is possible only in the presence of a theoretical
result that indicates how to compute the threshold that maximizes the
expected value of the effectiveness function \cite{Lewis95}. This is
typical of classifiers that output \emph{probability} estimates of the
membership of $d_j$ in $c_i$ (see Section \ref{sec:probabilistic}) and
whose effectiveness is computed by decision-theoretic measures such as
\emph{utility} (see Section \ref{sec:alternativestoeffectiveness}); we
thus defer the discussion of this policy (which is called {\em
probability thresholding} in \cite{Lewis95}) to Section
\ref{sec:alternativestoeffectiveness}.

When such a theoretical result is not available one has to revert to
the latter method, which consists in testing different values for
$\tau_i$ on a validation set and choosing the value which maximizes
effectiveness. We call this policy \emph{CSV thresholding}
\cite{Cohen99,Schapire98,Wiener95}; it is also called \emph{Scut} in
\cite{Yang99a}. Different $\tau_i$'s are typically chosen for the
different $c_i$'s.

A second, popular experimental policy is \emph{proportional
thresholding} \cite{Iwayama95,Larkey98,Lewis92,Lewis94,Wiener95}, also
called \emph{Pcut} in \cite{Yang99a}. This policy consists in choosing
the value of $\tau_i$ for which $g_{Va}(c_i)$ is closest to
$g_{Tr}(c_i)$, and embodies the principle that the same percentage of
documents of both training and test set should be classified under
$c_i$. For obvious reasons, this policy does not lend itself to
document-pivoted TC.

Sometimes, depending on the application, a \emph{fixed thresholding}
policy (aka ``$k$-per-doc'' thresholding \cite{Lewis92} or {\em Rcut}
\cite{Yang99a}) is applied, whereby it is stipulated that a fixed
number $k$ of categories, equal for all $d_j$'s, are to be assigned to
each document $d_j$. This is often used, for instance, in applications
of TC to automated document indexing \cite{Field75,Lam99a}. Strictly
speaking, however, this is not a thresholding policy in the sense
defined at the beginning of Section \ref{sec:construction}, as it
might happen that $d'$ is classified under $c_i$, $d''$ is not, and
$CSV_i(d')<CSV_i(d'')$. Quite clearly, this policy is mostly at home
with document-pivoted TC. However, it suffers from a certain
coarseness, as the fact that $k$ is equal for all documents (nor could
this be otherwise) allows no fine-tuning.

In his experiments Lewis \citeyear{Lewis92} found the proportional
policy to be superior to probability thresholding when microaveraged
effectiveness was tested but slightly inferior with macroaveraging
(see Section \ref{sec:precisionandrecall}). Yang \citeyear{Yang99a}
found instead $CSV$ thresholding to be superior to proportional
thresholding (possibly due to her category-specific optimization on a
validation set), and found fixed thresholding to be consistently
inferior to the other two policies. The fact that these latter results
have been obtained across different classifiers no doubt reinforce
them.

In general, aside from the considerations above, the choice of the
thresholding policy may also be influenced by the application; for
instance, in applying a text classifier to document indexing for
Boolean systems a fixed thresholding policy might be chosen, while a
proportional or $CSV$ thresholding method might be chosen for Web page
classification under hierarchical catalogues.


\subsection{Probabilistic classifiers}
\label{sec:probabilistic}

Probabilistic classifiers (see \cite{Lewis98} for a thorough
discussion) view $CSV_i(d_j)$ in terms of $P(c_i|\vec{d_j})$, i.e.\
the probability that a document represented by a vector
$\vec{d_j}=\langle w_{1j}, \ldots, w_{|{\cal T}| j}\rangle$ of (binary
or weighted) terms belongs to $c_i$, and compute this probability by
an application of Bayes' theorem, given by \begin{eqnarray}
\label{eq:bayestheorem} P(c_i|\vec{d_j}) & = & \frac{P(c_i)
P(\vec{d_j}|c_i)}{P(\vec{d_j})}\end{eqnarray}

\noindent In (\ref{eq:bayestheorem}) the event space is the space of
documents: $P(\vec{d_j})$ is thus the probability that a randomly
picked document has vector $\vec{d_j}$ as its representation, and
$P(c_i)$ the probability that a randomly picked document belongs to
$c_i$.

The estimation of $P(\vec{d_j}|c_i)$ in (\ref{eq:bayestheorem}) is
problematic, since the number of possible vectors $\vec{d_j}$ is too
high (the same holds for $P(\vec{d_j})$, but for reasons that will be
clear shortly this will not concern us). In order to alleviate this
problem it is common to make the assumption that any two coordinates
of the document vector are, when viewed as random variables,
statistically independent of each other; this \emph{independence
assumption} is encoded by the equation \begin{eqnarray}
\label{eq:independence} P(\vec{d_j}|c_i) & = & \prod_{k=1}^{|{\cal
T}|}P(w_{kj}|c_i)\end{eqnarray}

\noindent Probabilistic classifiers that use this assumption are
called \emph{Na\"{\i}ve Bayes} classifiers, and account for most of
the probabilistic approaches to TC in the literature (see e.g.\
\cite{Joachims98,Koller97,Larkey96,Lewis92,Lewis94a,Li98a,Robertson84}).
The ``na\"{\i}ve'' character of the classifier is due to the fact that
usually this assumption is, quite obviously, not verified in practice.

One of the best-known Na\"{\i}ve Bayes approaches is the \emph{binary
independence} classifier \cite{Robertson76}, which results from using
binary-valued vector representations for documents. In this case, if
we write $p_{ki}$ as short for $P(w_{kx}=1|c_i)$, the $P(w_{kj}|c_i)$
factors of (\ref{eq:independence}) may be written as \begin{eqnarray}
\label{eq:simplification} P(w_{kj}|c_i) & = & p_{ki}^{w_{kj}}
(1-p_{ki})^{1-w_{kj}} = (\frac{p_{ki}}{1-p_{ki}})^{w_{kj}} (1-p_{ki})
\end{eqnarray}

\noindent We may further observe that in TC the document space is
partitioned into two categories\footnote{Cooper \citeyear{Cooper95}
has pointed out that in this case the full independence assumption of
(\ref{eq:independence}) is not actually made in the Na\"{\i}ve Bayes
classifier; the assumption needed here is instead the weaker
\emph{linked dependence assumption}, which may be written as
$\frac{P(\vec{d_j}|c_i)}{P(\vec{d_j}|\overline{c}_i)}=\prod_{k=1}^{|{\cal
T}|} \frac{P(w_{kj}|c_i)}{P(w_{kj}|\overline{c}_i)}$.}, $c_i$ and its
complement $\overline{c}_i$, such that $P(\overline{c}_i|\vec{d_j}) =
1 - P(c_i|\vec{d_j})$. If we plug in (\ref{eq:independence}) and
(\ref{eq:simplification}) into (\ref{eq:bayestheorem}) and take logs
we obtain \begin{eqnarray} \label{eq:intermediate1} \log
P(c_i|\vec{d_j}) & = & \log P(c_i) + \\ & & \hspace{5ex}\nonumber
\sum_{k=1}^{|{\cal T}|}w_{kj} \log \frac{p_{ki}}{1-p_{ki}} +
\sum_{k=1}^{|{\cal T}|}\log (1-p_{ki}) -\log P(\vec{d_j}) \\
\label{eq:intermediate2} \log (1-P(c_i|\vec{d_j})) & = & \log
(1-P(c_i)) + \\ & & \hspace{5ex}\nonumber \sum_{k=1}^{|{\cal
T}|}w_{kj} \log \frac{p_{k\overline{i}}}{1-p_{k\overline{i}}} +
\sum_{k=1}^{|{\cal T}|}\log (1-p_{k\overline{i}}) -\log P(\vec{d_j})
\end{eqnarray}

\noindent where we write $p_{k\overline{i}}$ as short for
$P(w_{kx}=1|\overline{c}_i)$. We may convert (\ref{eq:intermediate1})
and (\ref{eq:intermediate2}) into a single equation by subtracting
componentwise (\ref{eq:intermediate2}) from (\ref{eq:intermediate1}),
thus obtaining \begin{eqnarray} \label{eq:final} \log
\frac{P(c_i|\vec{d_j})}{1-P(c_i|\vec{d_j})} & = & \log
\frac{P(c_i)}{1-P(c_i)} + \sum_{k=1}^{|{\cal T}|}w_{kj} \log
\frac{p_{ki}(1-p_{k\overline{i}})}{p_{k\overline{i}} (1-p_{ki})} +
\sum_{k=1}^{|{\cal T}|}\log \frac{1-p_{ki}}{1-p_{k\overline{i}}}
\end{eqnarray}

\noindent Note that $\frac{P(c_i|\vec{d_j})}{1-P(c_i|\vec{d_j})}$ is
an increasing monotonic function of $P(c_i|\vec{d_j})$, and may thus
be used directly as $CSV_i(d_j)$. Note also that $\log
\frac{P(c_i)}{1-P(c_i)}$ and $\sum_{k=1}^{|{\cal T}|}\log
\frac{1-p_{ki}}{1-p_{k\overline{i}}}$ are constant for all documents,
and may thus be disregarded\footnote{This is not true, however, if the
``fixed thresholding'' method of Section \ref{sec:thresholding} is
adopted. In fact, for a fixed document $d_j$ the first and third
factor in the formula above are different for different categories,
and may therefore influence the choice of the categories under which
to file $d_j$.}. Defining a classifier for category $c_i$ thus
basically requires estimating the $2|{\cal T}|$ parameters $\{p_{1i},
p_{1\overline{i}}, \ldots, p_{|{\cal T}| i}, p_{|{\cal
T}|\overline{i}}\}$ from the training data, which may be done in the
obvious way. Note that in general the classification of a given
document does not require to compute a sum of $|{\cal T}|$ factors, as
the presence of $\sum_{k=1}^{|{\cal T}|}w_{kj} \log
\frac{p_{ki}(1-p_{k\overline{i}})}{p_{k\overline{i}} (1-p_{ki})}$
would imply; in fact, all the factors for which $w_{kj}=0$ may be
disregarded, and this accounts for the vast majority of them, since
document vectors are usually very sparse.

The method we have illustrated is just one of the many variants of the
Na\"{\i}ve Bayes approach, the common denominator of which is
(\ref{eq:independence}). A recent paper by Lewis \citeyear{Lewis98} is
an excellent roadmap on the various directions that research on
Na\"{\i}ve Bayes classifiers has taken; among these are the ones
aiming
 
\begin{itemize}

\item \emph{to relax the constraint that document vectors should be
binary-valued}. This looks natural, given that weighted indexing
techniques (see e.g.\ \cite{Fuhr89,Salton88}) accounting for the
``importance'' of $t_k$ for $d_j$ play a key role in IR.

\item \emph{to introduce document length normalization}. The value of
$\log \frac{P(c_i|\vec{d_j})}{1-P(c_i|\vec{d_j})}$ tends to be more
extreme (i.e.\ very high or very low) for long documents (i.e.\
documents such that $w_{kj}=1$ for many values of $k$), irrespectively
of their semantic relatedness to $c_i$, thus calling for length
normalization. Taking length into account is easy in non-probabilistic
approaches to classification (see e.g.\ Section \ref{sec:Rocchio}),
but is problematic in probabilistic ones (see \cite[Section
5]{Lewis98}). One possible answer is to switch from an interpretation
of Na\"{\i}ve Bayes in which documents are events to one in which
terms are events \cite{Baker98,McCallum98b,Chakrabarti98c,Guthrie94}.
This accounts for document length naturally but, as noted in
\cite{Lewis98}, has the drawback that different occurrences of the
same word within the same document are viewed as independent, an
assumption even more implausible than (\ref{eq:independence}).

\item \emph{to relax the independence assumption}. This may be the
hardest route to follow, since this produces classifiers of higher
computational cost and characterized by harder parameter estimation
problems \cite{Koller97}. Earlier efforts in this direction within
probabilistic text search (e.g.\ \cite{vanRijsbergen77}) have not
shown the performance improvements that were hoped for. Recently, the
fact that the binary independence assumption seldom harms
effectiveness has also been given some theoretical justification
\cite{Domingos97}.

\end{itemize}

\noindent The quotation of text search in the last paragraph is not
casual. Unlike other types of classifiers, the literature on
probabilistic classifiers is inextricably intertwined with that on
probabilistic search systems (see \cite{Crestani98} for a review),
since these latter attempt to determine the probability that a
document falls in the category denoted by the query, and since they
are the only search systems that take \emph{relevance feedback}, a
notion essentially involving supervised learning, as central.


\subsection{Decision tree classifiers}
\label{sec:decisiontree}

Probabilistic methods are quantitative (i.e.\ numeric) in nature, and
as such have sometimes been criticized since, effective as they may
be, are not easily interpretable by humans. A class of algorithms that
do not suffer from this problem are \emph{symbolic} (i.e.\
non-numeric) algorithms, among which inductive rule learners (which we
will discuss in Section \ref{sec:rulelearning}) and decision tree
learners are the most important examples.

\begin{figure}[t]
\begin{center}
	\epsfig{file=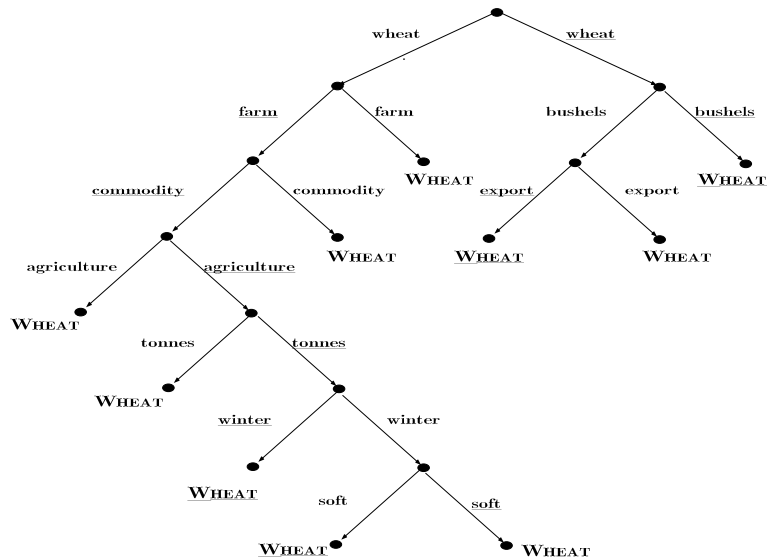, width=34em}
	\caption{\label{fig:decisiontree}A decision tree equivalent to
	the DNF rule of Figure \ref{fig:construe}. Edges are labelled
	by terms and leaves are labelled by categories
	(\underline{underlining} denotes negation).}\end{center}
\end{figure}

A \emph{decision tree} (DT) text classifier (see e.g.\ \cite[Chapter
3]{Mitchell96}) is a tree in which internal nodes are labelled by
terms, branches departing from them are labelled by tests on the
weight that the term has in the test document, and leafs are labelled
by categories. Such a classifier categorizes a test document $d_j$ by
recursively testing for the weights that the terms labeling the
internal nodes have in vector $\vec{d}_j$, until a leaf node is
reached; the label of this node is then assigned to $d_j$. Most such
classifiers use binary document representations, and thus consist of
binary trees. An example DT is illustrated in Figure
\ref{fig:decisiontree}.

There are a number of standard packages for DT learning, and most DT
approaches to TC have made use of one such package. Among the most
popular ones are ID3 (used in \cite{Fuhr91a}), C4.5 (used in
\cite{Cohen98,Cohen99,Joachims98,Lewis94c}) and C5 (used in
\cite{Li98a}). TC efforts based on experimental DT packages include
\cite{Dumais98,Lewis94,Weiss99}.

A possible method for learning a DT for category $c_i$ consists in a
``divide and conquer'' strategy of (i) checking whether all the
training examples have the same label (either $c_i$ or
$\overline{c}_i$); (ii) if not, selecting a term $t_k$, partitioning
$Tr$ into classes of documents that have the same value for $t_k$, and
placing each such class in a separate subtree. The process is
recursively repeated on the subtrees until each leaf of the tree so
generated contains training examples assigned to the same category
$c_i$, which is then chosen as the label for the leaf. The key step is
the choice of the term $t_k$ on which to operate the partition, a
choice which is generally made according to an information gain or
entropy criterion. However, such a ``fully grown'' tree may be prone
to overfitting, as some branches may be too specific to the training
data. Most DT learning methods thus include a method for growing the
tree and one for pruning it, i.e.\ for removing the overly specific
branches. Variations on this basic schema for DT learning abound
\cite[Section 3]{Mitchell96}.

DT text classifiers have been used either as the main classification
tool \cite{Fuhr91a,Lewis94c,Lewis94}, or as baseline classifiers
\cite{Cohen99,Joachims98}, or as members of classifier committees
\cite{Li98a,Schapire00,Weiss99}.


\subsection{Decision rule classifiers}
\label{sec:rulelearning}

A classifier for category $c_i$ built by an \emph{inductive rule
learning} method consists of a \emph{DNF rule}, i.e.\ of a conditional
rule with a premise in disjunctive normal form (DNF), of the type
illustrated in Figure \ref{fig:construe}\footnote{Many inductive rule
learning algorithms build \emph{decision lists} (i.e.\ arbitrarily
nested {\tt if-then-else} clauses) instead of DNF rules; since the
former may always be rewritten as the latter we will disregard the
issue.}. The literals (i.e.\ possibly negated keywords) in the premise
denote the presence (non-negated keyword) or absence (negated keyword)
of the keyword in the test document $d_{j}$, while the clause head
denotes the decision to classify $d_{j}$ under $c_i$. DNF rules are
similar to DTs in that they can encode any Boolean function. However,
an advantage of DNF rule learners is that they tend to generate more
compact classifiers than DT learners.

Rule learning methods usually attempt to select from all the possible
covering rules (i.e.\ rules that correctly classify all the training
examples) the ``best'' one according to some minimality criterion.
While DTs are typically built by a top-down, ``divide-and-conquer''
strategy, DNF rules are often built in a bottom-up fashion. Initially,
every training example $d_j$ is viewed as a clause $\eta_1, \ldots,
\eta_n \rightarrow \gamma_i$, where $\eta_1, \ldots, \eta_n$ are the
terms contained in $d_j$ and $\gamma_i$ equals $c_i$ or
$\overline{c}_i$ according to whether $d_j$ is a positive or negative
example of $c_i$. This set of clauses is already a DNF classifier for
$c_i$, but obviously scores high in terms of overfitting. The learner
applies then a process of generalization in which the rule is
simplified through a series of modifications (e.g.\ removing premises
from clauses, or merging clauses) that maximize its compactness while
at the same time not affecting the ``covering'' property of the
classifier. At the end of this process, a ``pruning'' phase similar in
spirit to that employed in DTs is applied, where the ability to
correctly classify \emph{all} the training examples is traded for more
generality.

DNF rule learners vary widely in terms of the methods, heuristics and
criteria employed for generalization and pruning. Among the DNF rule
learners that have been applied to TC are {\sc Charade}
\cite{Moulinier96a}, DL-ESC \cite{Li99}, {\sc Ripper}
\cite{Cohen95a,Cohen98,Cohen99}, {\sc Scar} \cite{Moulinier96}, and
\textsc{Swap-1} \cite{Apte94}.

While the methods above use rules of propositional logic (PL),
research has also been carried out using rules of first order logic
(FOL), obtainable through the use of \emph{inductive logic
programming} methods. Cohen \citeyear{Cohen95a} has extensively
compared PL and FOL learning in TC (for instance, comparing the PL
learner \textsc{Ripper} with its FOL version \textsc{Flipper}), and
has found that the additional representational power of FOL brings
about only modest benefits.


\subsection{Regression methods} \label{sec:regression}

Various TC efforts have used regression models (see e.g.\
\cite{Fuhr94,Ittner95,Lewis94a,Schutze95}). \emph{Regression} denotes
the approximation of a \emph{real-valued} (instead than binary, as in
the case of classification) function $\breve{\Phi}$ by means of a
function $\Phi$ that fits the training data \cite[page
236]{Mitchell96}. Here we will describe one such model, the {\em
Linear Least Squares Fit} (LLSF) applied to TC by Yang and Chute
\citeyear{Yang94}. In LLSF, each document $d_j$ has two vectors
associated to it: an \emph{input vector} $I(d_j)$ of $|{\cal T}|$
weighted terms, and an \emph{output vector} $O(d_j)$ of $|{\cal C}|$
weights representing the categories (the weights for this latter
vector are binary for training documents, and are non-binary $CSVs$
for test documents). Classification may thus be seen as the task of
determining an output vector $O(d_j)$ for test document $d_j$, given
its input vector $I(d_j)$; hence, building a classifier boils down to
computing a $|{\cal C}|\times |{\cal T}|$ matrix $\hat{M}$ such that
$\hat{M}I(d_j)=O(d_j)$.

LLSF computes the matrix from the training data by computing a linear
least-squares fit that minimizes the error on the training set
according to the formula $\hat{M} = \arg\min_{M} \| MI-O \|_F$, where
$\arg\min_{M}(x)$ stands as usual for the $M$ for which $x$ is
minimum, $\|V\|_F\stackrel{def}{=} \sqrt{\sum_{i=1}^{|{\cal
C}|}\sum_{j=1}^{|{\cal T}|}v_{ij}^{2}}$ represents the so-called {\em
Frobenius norm} of a $|{\cal C}|\times|{\cal T}|$ matrix, $I$ is the
$|{\cal T}|\times|Tr|$ matrix whose columns are the input vectors of
the training documents, and $O$ is the $|{\cal C}|\times |Tr|$ matrix
whose columns are the output vectors of the training documents. The
$\hat{M}$ matrix is usually computed by performing a singular value
decomposition on the training set, and its generic entry
$\hat{m}_{ik}$ represents the degree of association between category
$c_i$ and term $t_k$.

The experiments of \cite{Yang94,Yang99} indicate that LLSF is one of
the most effective text classifiers known to date. One of its
disadvantages, though, is that the cost of computing the $\hat{M}$
matrix is much higher than that of many other competitors in the TC
arena.


\subsection{On-line methods} \label{sec:online}

A \emph{linear classifier} for category $c_i$ is a vector
$\vec{c}_i=\langle w_{1i}, \ldots, w_{|{\cal T}|i}\rangle$ belonging
to the same $|{\cal T}|$-dimensional space in which documents are also
represented, and such that $CSV_i(d_j)$ corresponds to the dot product
$\sum_{k=1}^{|{\cal T}|}w_{ki}w_{kj}$ of $\vec{d}_j$ and $\vec{c}_i$.
Note that when both classifier and document weights are
cosine-normalized (see (\ref{eq:cosinenormalization})), the dot
product between the two vectors corresponds to their \emph{cosine
similarity}, i.e.\begin{eqnarray*} S(c_i,d_j) & = &
cos(\alpha)=\displaystyle\frac{\sum_{k=1}^{|{\cal T}|}w_{ki}\cdot
w_{jk}}{\sqrt{\sum_{k=1}^{|{\cal T}|}w_{ki}^2} \cdot
\sqrt{\sum_{k=1}^{|{\cal T}|}w_{kj}^2}} \end{eqnarray*}

\noindent which represents the cosine of the angle $\alpha$ that
separates the two vectors. This is the similarity measure between
query and document computed by standard vector-space IR engines, which
means in turn that once a linear classifier has been built,
classification can be performed by invoking such an engine.
Practically all search engines have a dot product flavour to them, and
can therefore be adapted to doing TC with a linear classifier.

Methods for learning linear classifiers are often partitioned in two
broad classes, batch methods and on-line methods.

\emph{Batch methods} build a classifier by analysing the training set
all at once. Within the TC literature, one example of a batch method
is \emph{linear discriminant analysis}, a model of the stochastic
dependence between terms that relies on the covariance matrices of the
categories \cite{Hull94,Schutze95}. However, the foremost example of a
batch method is the Rocchio method; because of its importance in the
TC literature this will be discussed separately in Section
\ref{sec:Rocchio}. In this section we will instead concentrate on
on-line classifiers.

\emph{On-line} (aka \emph{incremental}) \emph{methods} build a
classifier soon after examining the first training document, and
incrementally refine it as they examine new ones. This may be an
advantage in the applications in which $Tr$ is not available in its
entirety from the start, or in which the ``meaning'' of the category
may change in time, as e.g.\ in adaptive filtering. This is also apt
to applications (e.g.\ semi-automated classification, adaptive
filtering) in which we may expect the user of a classifier to provide
feedback on how test documents have been classified, as in this case
further training may be performed during the operating phase by
exploiting user feedback.

A simple on-line method is the \emph{perceptron} algorithm, first
applied to TC in \cite{Schutze95,Wiener95} and subsequently used in
\cite{Dagan97,Ng97}. In this algorithm, the classifier for $c_i$ is
first initialized by setting all weights $w_{ki}$ to the same positive
value. When a training example $d_j$ (represented by a vector
$\vec{d}_j$ of binary weights) is examined, the classifier built so
far classifies it. If the result of the classification is correct
nothing is done, while if it is wrong the weights of the classifier
are modified: if $d_j$ was a positive example of $c_i$ then the
weights $w_{ki}$ of ``active terms'' (i.e.\ the terms $t_k$ such that
$w_{kj}=1$) are ``promoted'' by increasing them by a fixed quantity
$\alpha>0$ (called \emph{learning rate}), while if $d_j$ was a
negative example of $c_i$ then the same weights are ``demoted'' by
decreasing them by $\alpha$. Note that when the classifier has reached
a reasonable level of effectiveness, the fact that a weight $w_{ki}$
is very low means that $t_k$ has negatively contributed to the
classification process so far, and may thus be discarded from the
representation. We may then see the perceptron algorithm (as all other
incremental learning methods) as allowing for a sort of ``on-the-fly
term space reduction'' \cite[Section 4.4]{Dagan97}. The perceptron
classifier has shown a good effectiveness in all the experiments
quoted above.

The perceptron is an \emph{additive weight-updating} algorithm. A
\emph{multiplicative} variant of it is \textsc{Positive Winnow}
\cite{Dagan97}, which differs from perceptron because two different
constants $\alpha_1>1$ and $0<\alpha_2<1$ are used for promoting and
demoting weights, respectively, and because promotion and demotion are
achieved by multiplying, instead of adding, by $\alpha_1$ and
$\alpha_2$. \emph{\textsc{Balanced Winnow}} \cite{Dagan97} is a
further variant of \textsc{Positive Winnow}, in which the classifier
consists of \emph{two} weights $w^+_{ki}$ and $w^-_{ki}$ for each term
$t_k$; the final weight $w_{ki}$ used in computing the dot product is
the difference $w^+_{ki}-w^-_{ki}$. Following the misclassification of
a positive instance, active terms have their $w^+_{ki}$ weight
promoted and their $w^-_{ki}$ weight demoted, whereas in the case of a
negative instance it is $w^+_{ki}$ that gets demoted while $w^-_{ki}$
gets promoted (for the rest, promotions and demotions are as in {\sc
Positive Winnow}). \textsc{Balanced Winnow} allows negative $w_{ki}$
weights, while in the perceptron and in \textsc{Positive Winnow} the
$w_{ki}$ weights are always positive. In experiments conducted by
Dagan et al.\ \citeyear{Dagan97}, \textsc{Positive Winnow} showed a
better effectiveness than perceptron but was in turn outperformed by
(Dagan et al.'s own version of) \textsc{Balanced Winnow}.

Other on-line methods for building text classifiers are {\sc
Widrow-Hoff}, a refinement of it called \textsc{Exponentiated
Gradient} (both applied for the first time to TC in \cite{Lewis96})
and {\sc Sleeping Experts} \cite{Cohen99}, a version of
\textsc{Balanced Winnow}. While the first is an additive
weight-updating algorithm, the second and third are multiplicative.
Key differences with the previously described algorithms are that
these three algorithms (i) update the classifier not only after
misclassifying a training example, but also after classifying it
correctly, and (ii) update the weights corresponding to all terms
(instead of just active ones).

Linear classifiers lend themselves to both category-pivoted and
document-pivoted TC. For the former the classifier $\vec{c}_i$ is
used, in a standard search engine, as a query against the set of test
documents, while for the latter the vector $\vec{d}_j$ representing
the test document is used as a query against the set of classifiers
$\{\vec{c}_1, \ldots, \vec{c}_{|{\cal C}|}\}$.


\subsection{The Rocchio method} \label{sec:Rocchio}

Some linear classifiers consist of an explicit \emph{profile} (or
prototypical document) of the category. This has obvious advantages in
terms of interpretability, as such a profile is more readily
understandable by a human than, say, a neural network classifier.
Learning a linear classifier is often preceded by local TSR; in this
case, a profile of $c_i$ is a weighted list of the terms whose
presence or absence is most useful for discriminating $c_i$.

The \emph{Rocchio method} is used for inducing linear, profile-style
classifiers. It relies on an adaptation to TC of the well-known
Rocchio's formula for relevance feedback in the vector-space model,
and it is perhaps the only TC method rooted in the IR tradition rather
than in the ML one. This adaptation was first proposed by Hull
\citeyear{Hull94}, and has been used by many authors since then,
either as an object of research in its own right
\cite{Ittner95,Joachims97,Sable00,Schapire98,Singhal97}, or as a
baseline classifier
\cite{Cohen99,Galavotti00,Joachims98,Lewis96,Schapire00,Schutze95}, or
as a member of a classifier committee \cite{Larkey96} (see Section
\ref{sec:committees}).

Rocchio's method computes a classifier $\vec{c}_i=\langle w_{1i},
\ldots, w_{|{\cal T}|i} \rangle$ for category $c_i$ by means of the
formula \begin{eqnarray} \nonumber w_{ki} & = & \beta\cdot\sum_{\{d_j
\in POS_i\}}\frac{w_{kj}}{|POS_i|} - \gamma\cdot\sum_{\{d_j \in
NEG_i\}}\frac{w_{kj}}{|NEG_i|}\end{eqnarray}

\noindent where $w_{kj}$ is the weight of $t_k$ in document $d_j$,
$POS_i=\{d_j \in Tr \ | \ \breve{\Phi}(d_j,c_i)=T \}$ and $NEG_i=\{d_j
\in Tr \ | \ \breve{\Phi}(d_j,c_i)=F \}$. In this formula, $\beta$ and
$\gamma$ are control parameters that allow setting the relative
importance of positive and negative examples. For instance, if $\beta$
is set to 1 and $\gamma$ to 0 (as e.g.\ in
\cite{Dumais98,Hull94,Joachims98,Schutze95}), the profile of $c_i$ is
the \emph{centroid} of its positive training examples. A classifier
built by means of the Rocchio method rewards the closeness of a test
document to the centroid of the positive training examples, and its
distance from the centroid of the negative training examples. The role
of negative examples is usually de-emphasized, by setting $\beta$ to a
high value and $\gamma$ to a low one (e.g.\ Cohen and Singer
\citeyear{Cohen99}, Ittner et al.\ \citeyear{Ittner95}, and Joachims
\citeyear{Joachims97} use $\beta=16$ and $\gamma=4$).

\begin{figure}[t]
\begin{center}\vspace{-1ex}
	\epsfig{file=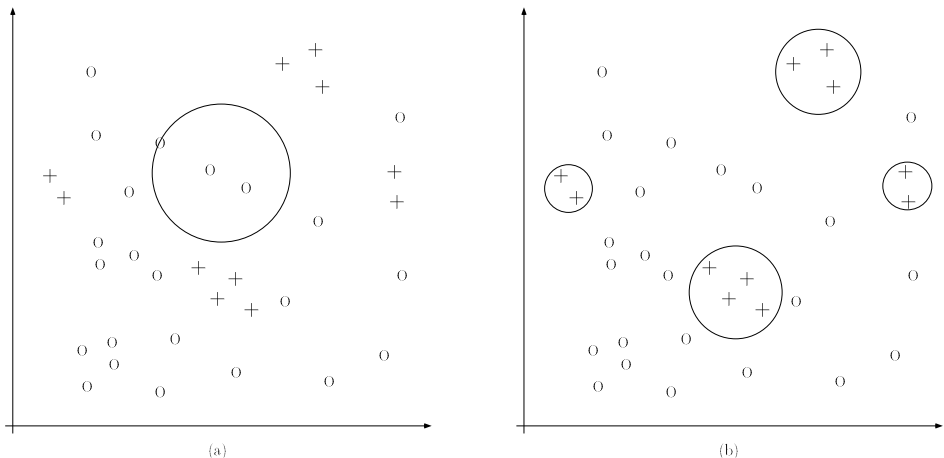, width=42em}
	\caption{\label{fig:rocchioKNN}A comparison between the TC
	behaviour of (a) the Rocchio classifier, and (b) the $k$-NN
	classifier. Small crosses and circles denote positive and
	negative training instances, respectively. The big circles
	denote the ``influence area'' of the classifier. Note that,
	for ease of illustration, document similarities are here
	viewed in terms of Euclidean distance rather than, as more
	common, in terms of dot product or cosine.}\end{center}
\end{figure}

This method is quite easy to implement, and is also quite efficient,
since learning a classifier basically comes down to averaging weights.
In terms of effectiveness, instead, a drawback is that if the
documents in the category tend to occur in disjoint clusters (e.g.\ a
set of newspaper articles lebelled with the {\tt Sports} category and
dealing with either boxing or rock-climbing), such a classifier may
miss most of them, as the centroid of these documents may fall outside
all of these clusters (see Figure \ref{fig:rocchioKNN}a). More
generally, a classifier built by the Rocchio method, as all linear
classifiers, has the disadvantage that it divides the space of
documents linearly. This situation is graphically depicted in Figure
\ref{fig:rocchioKNN}a, where documents are classified within $c_i$ if
and only if they fall within the circle. Note that even most of the
positive training examples would not be classified correctly by the
classifier.

\subsubsection{Enhancements to the basic Rocchio framework}
\label{sec:Rocchioenhancements}

One issue in the application of the Rocchio formula to profile
extraction is whether the set $NEG_i$ should be considered in its
entirety, or whether a well-chosen sample of it, such as the set
$NPOS_i$ of \emph{near-positives} (defined as ``the most positive
amongst the negative training examples''), should be selected from it,
yielding \begin{eqnarray} \nonumber w_{ki} & = & \beta\cdot\sum_{\{d_j
\in POS_i\}}\frac{w_{kj}}{|POS_i|} - \gamma\cdot\sum_{\{d_j \in
NPOS_i\}}\frac{w_{kj}}{|NPOS_i|}\end{eqnarray}

\noindent The $\sum_{\{d_j \in NPOS_i\}}\frac{w_{kj}}{|NPOS_i|}$
factor is more significant than $\sum_{\{d_j \in
NEG_i\}}\frac{w_{kj}}{|NEG_i|}$, since near-positives are the most
difficult documents to tell apart from the positives. Using
near-positives corresponds to the \emph{query zoning} method proposed
for IR by Singhal et al.\ \citeyear{Singhal97}. This method originates
from the observation that when the original Rocchio formula is used
for relevance feedback in IR, near-positives tend to be used rather
than generic negatives, as the documents on which user judgments are
available are among the ones that had scored highest in the previous
ranking. Early applications of the Rocchio formula to TC (e.g.
\cite{Hull94,Ittner95}) generally did not make a distinction between
near-positives and generic negatives. In order to select the
near-positives Schapire et al.\ \citeyear{Schapire98} issue a query,
consisting of the centroid of the positive training examples, against
a document base consisting of the negative training examples; the
top-ranked ones are the most similar to this centroid, and are then
the near-positives. Wiener et al.\ \citeyear{Wiener95} instead equate
the near-positives of $c_i$ to the positive examples of the
\emph{sibling} categories of $c_i$, as in the application they work on
(TC with hierarchical category sets) the notion of a ``sibling
category of $c_i$'' is well-defined. A similar policy is also adopted
in \cite{Ng97,Ruiz99,Weigend99}.

By using query zoning plus other enhancements (TSR, statistical
phrases, and a method called \emph{dynamic feedback optimization}),
Schapire et al.\ \citeyear{Schapire98} have found that a Rocchio
classifier can achieve an effectiveness comparable to that of a
state-of-the-art ML method such as ``boosting'' (see Section
\ref{sec:boosting}) while being 60 times quicker to train. These
recent results will no doubt bring about a renewed interest for the
Rocchio classifier, previously considered an underperformer
\cite{Cohen99,Joachims98,Lewis96,Schutze95,Yang99a}.


\subsection{Neural networks} \label{sec:neuralnetworks}

A \emph{neural network} (NN) text classifier is a network of units,
where the input units represent terms, the output unit(s) represent
the category or categories of interest, and the weights on the edges
connecting units represent dependence relations. For classifying a
test document $d_j$, its term weights $w_{kj}$ are loaded into the
input units; the activation of these units is propagated forward
through the network, and the value of the output unit(s) determines
the categorization decision(s). A typical way of training NNs is
backpropagation, whereby the term weights of a training document are
loaded into the input units, and if a misclassification occurs the
error is ``backpropagated'' so as to change the parameters of the
network and eliminate or minimize the error.

The simplest type of NN classifier is the perceptron
\cite{Dagan97,Ng97}, which is a linear classifier and as such has been
extensively discussed in Section \ref{sec:online}. Other types of
linear NN classifiers implementing a form of logistic regression have
also been proposed and tested by Sch\"utze et al.\
\citeyear{Schutze95} and Wiener et al.\ \citeyear{Wiener95}, yielding
very good effectiveness.

A non-linear NN
\cite{Lam99,Ruiz99,Schutze95,Weigend99,Wiener95,Yang99} is instead a
network with one or more additional ``layers'' of units, which in TC
usually represent higher-order interactions between terms that the
network is able to learn. When comparative experiments relating
non-linear NNs to their linear counterparts have been performed, the
former have yielded either no improvement \cite{Schutze95} or very
small improvements \cite{Wiener95} over the latter.


\subsection{Example-based classifiers}
\label{sec:examplebased}

Example-based classifiers do not build an explicit, declarative
representation of the category $c_i$, but rely on the category labels
attached to the training documents similar to the test document. These
methods have thus been called \emph{lazy learners}, since ``they defer
the decision on how to generalize beyond the training data until each
new query instance is encountered'' \cite[pag 244]{Mitchell96}.

The first application of example-based methods (aka \emph{memory-based
reasoning methods}) to TC is due to Creecy, Masand and colleagues
\cite{Creecy92,Masand92}; examples include
\cite{Joachims98,Lam99a,Larkey98,Larkey99,Li98a,Yang97,Yang99}. Our
presentation of the example-based approach will be based on the
\emph{$k$-NN} (for ``$k$ nearest neighbours'') algorithm used by Yang
\citeyear{Yang94a}. For deciding whether $d_j \in c_i$, $k$-NN looks
at whether the $k$ training documents most similar to $d_j$ also are
in $c_i$; if the answer is positive for a large enough proportion of
them, a positive decision is taken, and a negative decision is taken
otherwise. Actually, Yang's is a \emph{distance-weighted} version of
$k$-NN (see e.g.\ \cite[Section 8.2.1]{Mitchell96}), since the fact
that a most similar document is in $c_i$ is weighted by its similarity
with the test document. Classifying $d_{j}$ by means of $k$-NN thus
comes down to computing \begin{eqnarray} \label{eq:knn} CSV_i(d_j) & =
& \displaystyle\sum_{d_z \in \ Tr_k(d_j)}RSV(d_j,d_z)\cdot
\lsb\breve{\Phi}(d_z,c_i)\rsb
\end{eqnarray}

\noindent where $Tr_k(d_j)$ is the set of the $k$ documents $d_z$
which maximize $RSV(d_j,d_z)$ and $$ \lsb \alpha \rsb = \left \{
\begin{array}{rl} 1 & \mbox{if $\alpha=T$} \\ 0 & \mbox{if $\alpha=F$}
\end{array} \right . $$

\noindent The thresholding methods of Section \ref{sec:thresholding}
can then be used to convert the real-valued $CSV_i$'s into binary
categorization decisions. In (\ref{eq:knn}), $RSV(d_j,d_z)$ represents
some measure or semantic relatedness between a test document $d_j$ and
a training document $d_z$; any matching function, be it probabilistic
(as used in \cite{Larkey96}) or vector-based (as used in
\cite{Yang94a}), from a ranked IR system may be used for this purpose.
The construction of a $k$-NN classifier also involves determining
(experimentally, on a validation set) a threshold $k$ that indicates
how many top-ranked training documents have to be considered for
computing $CSV_i(d_j)$. Larkey and Croft \citeyear{Larkey96} use
$k=20$, while Yang \citeyear{Yang94a,Yang99a} has found $30\leq k\leq
45$ to yield the best effectiveness. Anyhow, various experiments have
shown that increasing the value of $k$ does not significantly degrade
the performance.

Note that $k$-NN, unlike linear classifiers, does not divide the
document space linearly, hence does not suffer from the problem
discussed at the end of Section \ref{sec:Rocchio}. This is graphically
depicted in Figure \ref{fig:rocchioKNN}b, where the more ``local''
character of $k$-NN with respect to Rocchio can be appreciated.

This method is naturally geared towards document-pivoted TC, since
ranking the training documents for their similarity with the test
document can be done once for all categories. For category-pivoted TC
one would need to store the document ranks for each test document,
which is obviously clumsy; DPC is thus {\it de facto} the only
reasonable way to use $k$-NN.

A number of different experiments (see Section \ref{sec:best}) have
shown $k$-NN to be quite effective. However, its most important
drawback is its inefficiency at classification time: while e.g.\ with
a linear classifier only a dot product needs to be computed to
classify a test document, $k$-NN requires the entire training set to
be ranked for similarity with the test document, which is much more
expensive. This is a drawback of ``lazy'' learning methods, since they
do not have a true training phase and thus defer all the computation
to classification time.


\subsubsection{Other example-based techniques}
\label{sec:otherexamplebased}

Various example-based techniques have been used in the TC literature.
For example, Cohen and Hirsh \citeyear{Cohen98} implement an
example-based classifier by extending standard relational DBMS
technology with ``similarity-based soft joins''. In their
\textsc{Whirl} system they use the scoring function \begin{eqnarray*}
\label{eq:Cohensknn} CSV_i(d_j) & = & 1-\displaystyle\prod_{d_z \in \
Tr_k(d_j)}(1-RSV(d_j,d_z))^{\lsb\breve{\Phi}(d_z,c_i)\rsb}
\end{eqnarray*}

\noindent as an alternative to (\ref{eq:knn}), obtaining a small but
statistically significant improvement over a version of \textsc{Whirl}
using (\ref{eq:knn}). In their experiments this technique outperformed
a number of other classifiers, such as a C4.5 decision tree classifier
and the \textsc{Ripper} CNF rule-based classifier.
 
A variant of the basic $k$-NN approach is proposed by Galavotti et al.
\citeyear{Galavotti00}, who reinterpret (\ref{eq:knn}) by redefining
$\lsb\alpha\rsb$ as $$ \lsb \alpha \rsb = \left \{ \begin{array}{rl} 1
& \mbox{if $\alpha=T$} \\ -1 & \mbox{if $\alpha=F$} \end{array} \right
. $$

\noindent The difference from the original $k$-NN approach is that if
a training document $d_z$ similar to the test document $d_j$ does not
belong to $c_i$, this information is not discarded but weights
\emph{negatively} in the decision to classify $d_j$ under $c_i$.

A combination of profile- and example-based methods is presented in
\cite{Lam98}. In this work a $k$-NN system is fed \emph{generalized
instances} (GIs) in place of training documents. This approach may be
seen as the result of

\begin{itemize}

\item clustering the training set, thus obtaining a set of clusters
$K_{i}=\{k_{i1}, \ldots, k_{i|K_{i}|}\}$;

\item building a profile $G(k_{iz})$ (``generalized instance'') from
the documents belonging to cluster $k_{iz}$ by means of some algorithm
for learning linear classifiers (e.g.\ Rocchio, \textsc{Widrow-Hoff});

\item applying $k$-NN with profiles in place of training documents,
i.e.\ computing \begin{eqnarray} \nonumber CSV_i(d_j) &
\stackrel{def}{=} & \displaystyle\sum_{k_{iz}\in
K_{i}}RSV(d_j,G(k_{iz}))\cdot \frac{|\{d_j \in k_{iz} | \
\breve{\Phi}(d_j,c_i)=T \}|}{|\{d_j \in k_{iz} \}|}\cdot \frac{|\{d_j
\in k_{iz} \}|}{|Tr|} \\ & = & \displaystyle\sum_{k_{iz}\in
K_{i}}RSV(d_j,G(k_{iz}))\cdot \frac{|\{d_j \in k_{iz} | \
\breve{\Phi}(d_j,c_i)=T \}|}{|Tr|}
\end{eqnarray}

\noindent where $\frac{|\{d_j \in k_{iz} | \ \breve{\Phi}(d_j,c_i)=T
\}|}{|\{d_j \in k_{iz} \}|}$ represents the ``degree'' to which
$G(k_{iz})$ is a positive instance of $c_i$, and $\frac{|\{d_j \in
k_{iz} \}|}{|Tr|}$ represents its weight within the entire process.

\end{itemize}

\noindent This exploits the superior effectiveness (see Figure
\ref{fig:rocchioKNN}) of $k$-NN over linear classifiers while at the
same time avoiding the sensitivity of $k$-NN to the presence of
``outliers'' (i.e.\ positive instances of $c_i$ that ``lie out'' of
the region where most other positive instances of $c_i$ are located)
in the training set.


\subsection{Building classifiers by support vector machines}
\label{sec:supportvectormachines}

The \emph{support vector machine} (SVM) method has been introduced in
TC by Joachims \citeyear{Joachims98,Joachims99} and subsequently used
in \cite{Drucker99,Dumais98,Dumais00,Klinkenberg00,Taira99,Yang99}. In
geometrical terms, it may be seen as the attempt to find, among all
the surfaces $\sigma_1, \sigma_2, \ldots$ in $|{\cal T}|$-dimensional
space that separate the positive from the negative training examples
(\emph{decision surfaces}), the $\sigma_i$ that separates the
positives from the negatives by the widest possible margin, i.e.\ such
that the separation property is invariant with respect to the widest
possible traslation of $\sigma_i$.
 
\begin{figure}[t]
\begin{center}\vspace{-1ex}
\epsfig{file=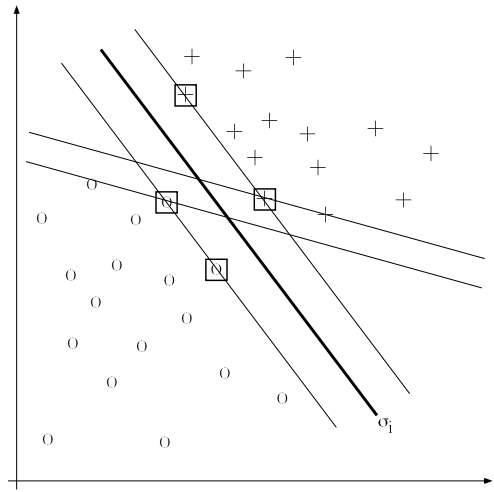, width=22em}
\caption{\label{fig:supportvector}Learning support vector classifiers.
The small crosses and circles represent positive and negative training
examples, respectively, whereas lines represent decision surfaces.
Decision surface $\sigma_i$ (indicated by the thicker line) is, among
those shown, the best possible one, as it is the middle element of the
widest set of parallel decision surfaces (i.e.\ its minimum distance
to any training example is maximum). Small boxes indicate the support
vectors.}\end{center}
\end{figure}
 
This idea is best understood in the case in which the positives and
the negatives are linearly separable, in which case the decision
surfaces are $(|{\cal T}|-1)$-hyperplanes. In the 2-dimensional case
of Figure \ref{fig:supportvector}, various lines may be chosen as
decision surfaces. The SVM method chooses the middle element from the
``widest'' set of parallel lines, i.e.\ from the set in which the
maximum distance between two elements in the set is highest. It is
noteworthy that this ``best'' decision surface is determined by only a
small set of training examples, called the \emph{support vectors}.

The method described is applicable also to the case in which the
positives and the negatives are not linearly separable. Yang and Liu
\citeyear{Yang99} experimentally compared the linear case (namely,
when the assumption is made that the categories are linearly
separable) with the non-linear case on a standard benchmark, and
obtained slightly better results in the former case.

As argued by Joachims \citeyear{Joachims98}, SVMs offer two important
advantages for TC:

\begin{itemize}

\item term selection is often not needed, as SVMs tend to be fairly
robust to overfitting and can scale up to considerable
dimensionalities;

\item no human and machine effort in parameter tuning on a validation
set is needed, as there is a theoretically motivated, ``default''
choice of parameter settings, which has also been shown to provide the
best effectiveness.

\end{itemize}

\noindent Dumais et al.\ \citeyear{Dumais98} have applied a novel
algorithm for training SVMs which brings about training speeds
comparable to computationally easy learners such as Rocchio.


\subsection{Classifier committees} \label{sec:committees}

Classifier \emph{committees} (aka \emph{ensembles}) are based on the
idea that, given a task that requires expert knowledge to perform, $k$
experts may be better than one if their individual judgments are
appropriately combined. In TC, the idea is to apply $k$ different
classifiers $\Phi_1, \ldots, \Phi_k$ to the same task of deciding
whether $d_j \in c_i$, and then combine their outcome appropriately. A
classifier committee is then characterized by (i) a choice of $k$
classifiers, and (ii) a choice of a combination function.

Concerning issue (i), it is known from the ML literature that, in
order to guarantee good effectiveness, the classifiers forming the
committee should be as independent as possible \cite{Tumer96}. The
classifiers may differ for the indexing approach used, or for the
inductive method, or both. Within TC, the avenue which has been
explored most is the latter (to our knowledge the only example of the
former is \cite{Scott99}).

Concerning issue (ii), various rules have been tested. The simplest
one is \emph{majority voting} (MV), whereby the binary outputs of the
$k$ classifiers are pooled together, and the classification decision
that reaches the majority of $\frac{k+1}{2}$ votes is taken ($k$
obviously needs to be an odd number) \cite{Li98a,Liere97}. This method
is particularly suited to the case in which the committee includes
classifiers characterized by a binary decision function $CSV_i:{\cal
D}\rightarrow \{T,F\}$. A second rule is \emph{weighted linear
combination} (WLC), whereby a weighted sum of the $CSV_i$'s produced
by the $k$ classifiers yields the final $CSV_i$. The weights $w_j$
reflect the expected relative effectiveness of classifiers $\Phi_j$,
and are usually optimized on a validation set \cite{Larkey96}. Another
policy is \emph{dynamic classifier selection} (DCS), whereby among
committee \{$\Phi_1$, \ldots, $\Phi_k$\} the classifier $\Phi_t$ most
effective on the $l$ validation examples most similar to $d_j$ is
selected, and its judgment adopted by the committee \cite{Li98a}. A
still different policy, somehow intermediate between WLC and DCS, is
\emph{adaptive classifier combination} (ACC), whereby the judgments of
\emph{all} the classifiers in the committee are summed together, but
their individual contribution is weighted by their effectiveness on
the $l$ validation examples most similar to $d_j$ \cite{Li98a}.

Classifier committees have had mixed results in TC so far. Larkey and
Croft \citeyear{Larkey96} have used combinations of Rocchio,
Na\"{\i}ve Bayes and $k$-NN, all together or in pairwise combinations,
using a WLC rule. In their experiments the combination of any two
classifiers outperformed the best individual classifier ($k$-NN), and
the combination of the three classifiers improved an all three
pairwise combinations. These results would seem to give strong support
to the idea that classifier committees can somehow profit from the
complementary strengths of their individual members. However, the
small size of the test set used (187 documents) suggests that more
experimentation is needed before conclusions can be drawn.

Li and Jain \citeyear{Li98a} have tested a committee formed of
(various combinations of) a Na\"{\i}ve Bayes classifier, an
example-based classifier, a decision tree classifier, and a classifier
built by means of their own ``subspace method''; the combination rules
they have worked with are MV, DCS and ACC. Only in the case of a
committee formed by Na\"{\i}ve Bayes and the subspace classifier
combined by means of ACC the committee has outperformed, and by a
narrow margin, the best individual classifier (for every attempted
classifier combination ACC gave better results than MV and DCS). This
seems discouraging, especially in the light of the fact that the
committee approach is computationally expensive (its cost trivially
amounts to the sum of the costs of the individual classifiers plus the
cost incurred for the computation of the combination rule). Again, it
has to be remarked that the small size of their experiment (two test
sets of less than 700 documents each were used) does not allow to draw
definitive conclusions on the approaches adopted.


\subsubsection{Boosting} \label{sec:boosting}

The \emph{boosting} method \cite{Schapire98,Schapire00} occupies a
special place in the classifier committees literature, since the $k$
classifiers $\Phi_1, \ldots, \Phi_{k}$ forming the committee are
obtained by the \emph{same} learning method (here called the
\emph{weak learner}). The key intuition of boosting is that the $k$
classifiers should be trained not in a conceptually parallel and
independent way, as in the committees described above, but
sequentially. In this way, in training classifier $\Phi_i$ one may
take into account how classifiers $\Phi_1, \ldots, \Phi_{i-1}$ perform
on the training examples, and concentrate on getting right those
examples on which $\Phi_1, \ldots, \Phi_{i-1}$ have performed worst.
 
Specifically, for learning classifier $\Phi_t$ each $\langle
d_j,c_i\rangle$ pair is given an ``importance weight'' $h^{t}_{ij}$
(where $h^{1}_{ij}$ is set to be equal for all $\langle
d_j,c_i\rangle$ pairs\footnote{Schapire et al.\ \citeyear{Schapire98}
also show that a simple modification of this policy allows
optimization of the classifier based on ``utility'' (see Section
\ref{sec:alternativestoeffectiveness}).}), which represents how hard
to get a correct decision for this pair was for classifiers $\Phi_1,
\ldots, \Phi_{t-1}$. These weights are exploited in learning $\Phi_t$,
which will be specially tuned to correctly solve the pairs with higher
weight. Classifier $\Phi_t$ is then applied to the training documents,
and as a result weights $h^{t}_{ij}$ are updated to $h^{t+1}_{ij}$; in
this update operation, pairs correctly classified by $\Phi_t$ will
have their weight decreased, while pairs misclassified by $\Phi_t$
will have their weight increased. After all the $k$ classifiers have
been built, a weighted linear combination rule is applied to yield the
final committee.

In the \textsc{BoosTexter} system \cite{Schapire00}, two different
boosting algorithms are tested, using a one-level decision tree weak
learner. The former algorithm (\textsc{AdaBoost.MH}, simply called
{\sc AdaBoost} in \cite{Schapire98}) is explicitly geared towards the
maximization of microaveraged effectiveness, whereas the latter
(\textsc{AdaBoost.MR}) is aimed at minimizing \emph{ranking loss}
(i.e.\ at getting a correct category ranking for each individual
document). In experiments conducted over three different test
collections, Schapire et al.\ \citeyear{Schapire98} have shown {\sc
AdaBoost} to outperform \textsc{Sleeping Experts}, a classifier that
had proven quite effective in the experiments of \cite{Cohen99}.
Further experiments by Schapire and Singer \citeyear{Schapire00}
showed {\sc AdaBoost} to outperform, aside from \textsc{Sleeping
Experts}, a Na\"{\i}ve Bayes classifier, a standard (non-enhanced)
Rocchio classifier, and Joachims' \citeyear{Joachims97}
\textsc{PrTFIDF} classifier.

A boosting algorithm based on a ``committee of classifier
sub-committees'' that improves on the effectiveness and (especially)
the efficiency of \textsc{AdaBoost.MH} is presented in
\cite{Sebastiani00}. An approach similar to boosting is also employed
by Weiss et al.\ \citeyear{Weiss99}, who experiment with committees of
decision trees each having an average of 16 leaves (hence much more
complex than the simple 2-leaves trees used in \cite{Schapire00}),
eventually combined by using the simple MV rule as a combination rule;
similarly to boosting, a mechanism for emphasising documents that have
been misclassified by previous decision trees is used. Boosting-based
approaches have also been employed in
\cite{Escudero00,Iyer00,Kim00,Li98a,Myers00}.


\subsection{Other methods} \label{sec:other}

Although in the previous sections we have tried to give an overview as
complete as possible of the learning approaches proposed in the TC
literature, it would be hardly possible to be exhaustive. Some of the
learning approaches adopted do not fall squarely under one or the
other class of algorithms, or have remained somehow isolated attempts.
Among these, the most noteworthy are the ones based on \emph{Bayesian
inference networks} \cite{Dumais98,Lam97,Tzeras93}, \emph{genetic
algorithms} \cite{Clack97,Masand94}, and \emph{maximum entropy
modelling} \cite{Manning99}.


\section{Evaluation of text classifiers}
\label{sec:evaluation}

As for text search systems, the evaluation of document classifiers is
typically conducted \emph{experimentally}, rather than analytically.
The reason is that, in order to evaluate a system analytically (e.g.\
proving that the system is correct and complete) we would need a
formal specification of the problem that the system is trying to solve
(e.g.\ \emph{with respect to what} correctness and completeness are
defined), and the central notion of TC (namely, that of membership of
a document in a category) is, due to its subjective character,
inherently non-formalisable.

The experimental evaluation of a classifier usually measures its
\emph{effectiveness} (rather than its efficiency), i.e.\ its ability
to take the \emph{right} classification decisions.
 

\subsection{Measures of text categorization effectiveness}
\label{sec:effectivenessmeasures}


\subsubsection{Precision and recall}
\label{sec:precisionandrecall}

Classification effectiveness is usually measured in terms of the
classic IR notions of precision ($\pi$) and recall ($\rho$), adapted
to the case of TC. \emph{Precision wrt $c_i$} ($\pi_i$) is defined as
the conditional probability $P(\breve{\Phi}(d_x,c_i)=T \ |
\Phi(d_x,c_i)=T)$, i.e.\ as the probability that if a random document
$d_x$ is classified under $c_i$, this decision is correct.
Analogously, \emph{recall wrt $c_i$} ($\rho_i$) is defined as
$P(\Phi(d_x,c_i)=T \ | \ \breve{\Phi}(d_x,c_i)=T)$, i.e.\ as the
probability that, if a random document $d_x$ ought to be classified
under $c_i$, this decision is taken. These category-relative values
may be averaged, in a way to be discussed shortly, to obtain $\pi$ and
$\rho$, i.e.\ values global to the entire category set. Borrowing
terminology from logic, $\pi$ may be viewed as the ``degree of
soundness'' of the classifier wrt ${\cal C}$, while $\rho$ may be
viewed as its ``degree of completeness'' wrt ${\cal C}$. As defined
here, $\pi_i$ and $\rho_i$ are to be understood as \emph{subjective}
probabilities, i.e.\ as measuring the expectation of the user that the
system will behave correctly when classifying an unseen document under
$c_i$. These probabilities may be estimated in terms of the
\emph{contingency table} for $c_i$ on a given test set (see Table
\ref{tab:contingency}). Here, $FP_i$ (\emph{false positives wrt
$c_i$}, aka \emph{errors of commission}) is the number of test
documents incorrectly classified under $c_i$; $TN_i$ (\emph{true
negatives wrt $c_i$}), $TP_i$ (\emph{true positives wrt $c_i$}) and
$FN_i$ (\emph{false negatives wrt $c_i$}, aka \emph{errors of
omission}) are defined accordingly. Estimates (indicated by carets) of
precision wrt $c_i$ and recall wrt $c_i$ may thus be obtained as
\begin{eqnarray*}\hat{\pi}_i = \displaystyle\frac{TP_i}{TP_i + FP_i}
\hspace{5em} \hat{\rho}_i = \displaystyle\frac{TP_i}{TP_i + FN_i}
\end{eqnarray*}

\begin{table}\begin{center}\begin{tabular}{|c|c||c|c|}
\hline \multicolumn{2}{|c||}{Category} & \multicolumn{2}{c|}{expert
judgments} \\ \cline{3-4} \multicolumn{2}{|c||}{$c_i$} & \textbf{YES}
& \textbf{NO} \\
\hline \hline classifier & \textbf{YES} & $TP_i$ & $FP_i$ \\
\cline{2-4} judgments & \textbf{NO} & $FN_i$ & $TN_i$ \\ \hline
\end{tabular}\caption{\label{tab:contingency}The contingency table for
category $c_i$.}\end{center}\end{table}

\noindent For obtaining estimates of $\pi$ and $\rho$, two different
methods may be adopted:

\begin{itemize}

\item \emph{microaveraging}: $\pi$ and $\rho$ are obtained by summing
over all individual decisions: \begin{eqnarray*} \hat{\pi}^\mu & = &
\displaystyle\frac{TP}{TP + FP} =
\displaystyle\frac{\sum_{i=1}^{|{\cal C}|} TP_i}{\sum_{i=1}^{|{\cal
C}|} (TP_i + FP_i)} \\ \hat{\rho}^\mu & = & \displaystyle\frac{TP}{TP
+ FN} = \displaystyle\frac{\sum_{i=1}^{|{\cal C}|}
TP_i}{\sum_{i=1}^{|{\cal C}|} (TP_i + FN_i)}\end{eqnarray*}

where ``$\mu$'' indicates microaveraging. The ``global'' contingency
table (Table \ref{tab:contingencyglobal}) is thus obtained by summing
over category-specific contingency tables.

\begin{table}\begin{center}\begin{tabular}{|c|c||c|c|}
\hline \multicolumn{2}{|c||}{Category set} &
\multicolumn{2}{c|}{expert judgments} \\ \cline{3-4}
\multicolumn{2}{|c||}{${\cal C}=\{c_1, \ldots, c_{|{\cal C}|}\}$} &
\textbf{YES} & \textbf{NO} \\ \hline \hline classifier & \textbf{YES}
& $TP=\displaystyle\sum_{i=1}^{|{\cal C}|} TP_i$ &
$FP=\displaystyle\sum_{i=1}^{|{\cal C}|} FP_i$ \\ \cline{2-4}
judgments & \textbf{NO} & $FN=\displaystyle\sum_{i=1}^{|{\cal C}|}
FN_i$ & $TN=\displaystyle\sum_{i=1}^{|{\cal C}|} TN_i$ \\ \hline
\end{tabular}\caption{\label{tab:contingencyglobal}The global
contingency table.}\end{center}\end{table}

\item \emph{macroaveraging} : precision and recall are first evaluated
``locally'' for each category, and then ``globally'' by averaging over
the results of the different categories: \begin{eqnarray*} \hat{\pi}^M
= \frac{\sum_{i=1}^{|{\cal C}|}\hat{\pi}_i}{|{\cal C}|} \hspace{5em}
\hat{\rho}^M = \frac{\sum_{i=1}^{|{\cal C}|}\hat{\rho}_i}{|{\cal C}|}
\end{eqnarray*}

where ``$M$'' indicates macroaveraging.

\end{itemize}

\noindent These two methods may give quite different results,
especially if the different categories have very different generality.
For instance, the ability of a classifier to behave well also on
categories with low generality (i.e.\ categories with few positive
training instances) will be emphasized by macroaveraging and much less
so by microaveraging. Whether one or the other should be used
obviously depends on the application requirements. From now on, we
will assume that microaveraging is used; everything we will say in the
rest of Section \ref{sec:evaluation} may be adapted to the case of
macroaveraging in the obvious way.


\subsubsection{Other measures of effectiveness}
\label{sec:othereffectiveness}

Measures alternative to $\pi$ and $\rho$ and commonly used in the ML
literature, such as \emph{accuracy} (estimated as $\hat{A} =
\frac{TP+TN}{TP+TN+FP+FN}$) and \emph{error} (estimated as
$\hat{E}=\frac{FP+FN}{TP+TN+FP+FN}=1-\hat{A}$), are not widely used in
TC. The reason is that, as Yang \citeyear{Yang99a} points out, the
large value that their denominator typically has in TC makes them much
more insensitive to variations in the number of correct decisions ($TP
+ TN$) than $\pi$ and $\rho$. Besides, if $A$ is the adopted
evaluation measure, in the frequent case of a very low average
generality the \emph{trivial rejector} (i.e.\ the classifier $\Phi$
such that $\Phi(d_j,c_i)=F$ for all $d_j$ and $c_i$) tends to
outperform all non-trivial classifiers (see also \cite[Section
2.3]{Cohen95a}). If $A$ is adopted, parameter tuning on a validation
set may thus result in parameter choices that make the classifier
behave very much like the trivial rejector.

A non-standard effectiveness measure is proposed by Sable and
Hatzivassiloglou \citeyear[Section 7]{Sable00}, who suggest to base
$\pi$ and $\rho$ not on ``absolute'' values of success and failure
(i.e.\ 1 if $\Phi(d_j,c_i)=\breve{\Phi}(d_j,c_i)$ and 0 if
$\Phi(d_j,c_i)\not =\breve{\Phi}(d_j,c_i)$), but on values of
\emph{relative success} (i.e.\ $CSV_i(d_j)$ if
$\breve{\Phi}(d_j,c_i)=T$ and $1-CSV_i(d_j)$ if
$\breve{\Phi}(d_j,c_i)=F$). This means that for a correct (resp.\
wrong) decision the classifier is rewarded (resp.\ penalized)
proportionally to its confidence in the decision. This proposed
measure does not reward the choice of a good thresholding policy, and
is thus unfit for autonomous (``hard'') classification systems.
However, it might be appropriate for interactive (``ranking'')
classifiers of the type used in \cite{Larkey99}, where the confidence
that the classifier has in its own decision influences category
ranking and, as a consequence, the overall usefulness of the system.


\subsubsection{Measures alternative to effectiveness}
\label{sec:alternativestoeffectiveness}

In general, criteria different from effectiveness are seldom used in
classifier evaluation. For instance, \emph{efficiency}, although very
important for applicative purposes, is seldom used as the sole
yardstick, due to the volatility of the parameters on which the
evaluation rests. However, efficiency may be useful for choosing among
classifiers with similar effectiveness. An interesting evaluation has
been carried out by Dumais et al.\ \citeyear{Dumais98}, who have
compared five different learning methods along three different
dimensions, namely effectiveness, \emph{training efficiency} (i.e.\
the average time it takes to build a classifier for category $c_i$
from a training set $Tr$), and \emph{classification efficiency} (i.e.\
the average time it takes to classify a new document $d_j$ under
category $c_i$).

An important alternative to effectiveness is \emph{utility}, a class
of measures from decision theory that extend effectiveness by economic
criteria such as \emph{gain} or \emph{loss}. Utility is based on a
{\em utility matrix} such as that of Table \ref{tab:utilitymatrix},
where the numeric values $u_{TP}$, $u_{FP}$, $u_{FN}$ and $u_{TN}$
represent the gain brought about by a true positive, false positive,
false negative and true negative, respectively; both $u_{TP}$ and
$u_{TN}$ are greater than both $u_{FP}$ and $u_{FN}$. ``Standard''
effectiveness is a special case of utility, i.e.\ the one in which
$u_{TP}=u_{TN}>u_{FP}=u_{FN}$. Less trivial cases are those in which
$u_{TP}\not =u_{TN}$ and/or $u_{FP}\not =u_{FN}$; this is the case
e.g.\ in spam filtering, where failing to discard a piece of junk mail
(FP) is a less serious mistake than discarding a legitimate message
(FN) \cite{Androutsopoulos00}. If the classifier outputs probability
estimates of the membership of $d_j$ in $c_i$, then decision theory
provides \emph{analytical} methods to determine thresholds $\tau_i$,
thus avoiding the need to determine them experimentally (as discussed
in Section \ref{sec:thresholding}). Specifically, as Lewis
\citeyear{Lewis95} reminds, the expected value of utility is maximized
when \begin{eqnarray}\nonumber\tau_i & = & \frac{(u_{FP}-
u_{TN})}{(u_{FN}- u_{TP}) + (u_{FP}- u_{TN})}\
\end{eqnarray}

\noindent which, in the case of ``standard'' effectiveness, is equal
to $\frac{1}{2}$.

The use of utility in TC is discussed in detail by Lewis
\citeyear{Lewis95}. Other works where utility is employed are
\cite{Amati99,Cohen99,Hull96,Lewis94c,Schapire98}. Utility has become
popular within the text filtering community, and the TREC ``filtering
track'' evaluations have been using it since long \cite{Lewis95b}. The
values of the utility matrix are extremely application-dependent. This
means that if utility is used instead of ``pure'' effectiveness, there
is a further element of difficulty in the cross-comparison of
classification systems (see Section \ref{sec:best}), since for two
classifiers to be experimentally comparable also the two utility
matrices must be the same.

\begin{table}\begin{center}\begin{tabular}{|c|c||c|c|}
\hline \multicolumn{2}{|c||}{Category set} &
\multicolumn{2}{c|}{expert judgments} \\ \cline{3-4}
\multicolumn{2}{|c||}{${\cal C}=\{c_1, \ldots, c_{|{\cal C}|}\}$} &
\textbf{YES} & \textbf{NO} \\ \hline \hline classifier & \textbf{YES}
& $u_{TP}$ & $u_{FP}$ \\ \cline{2-4} judgments & \textbf{NO} &
$u_{FN}$ & $u_{TN}$ \\ \hline
\end{tabular}\caption{\label{tab:utilitymatrix}The utility
matrix.}\end{center}\end{table}

Other effectiveness measures different from the ones discussed here
have occasionally been used in the literature; these include
\emph{adjacent score} \cite{Larkey98}, \emph{coverage}
\cite{Schapire00}, \emph{one-error} \cite{Schapire00}, \emph{Pearson
product-moment correlation} \cite{Larkey98}, \emph{recall at $n$}
\cite{Larkey96}, \emph{top candidate} \cite{Larkey96}, \emph{top $n$}
\cite{Larkey96}. We will not attempt to discuss them in detail.
However, their use shows that, although the TC community is making
consistent efforts at standardising experimentation protocols, we are
still far from universal agreement on evaluation issues and, as a
consequence, from understanding precisely the relative merits of the
various methods.


\subsubsection{Combined effectiveness measures}
\label{sec:combinedmeasures}

Neither precision nor recall make sense in isolation of each other. In
fact the classifier $\Phi$ such that $\Phi(d_j,c_i)=T$ for all $d_j$
and $c_i$ (the \emph{trivial acceptor}) has $\rho=1$. When the $CSV_i$
function has values in $[0,1]$ one only needs to set every threshold
$\tau_i$ to $0$ to obtain the trivial acceptor. In this case $\pi$
would usually be very low (more precisely, equal to the average test
set generality $\frac{\sum_{i=1}^{|{\cal C}|}g_{Te}(c_i)}{|{\cal
C}|}$)\footnote{From this one might be tempted to infer, by symmetry,
that the trivial rejector always has $\pi=1$. This is false, as $\pi$
is undefined (the denominator is zero) for the trivial rejector (see
Table \ref{tab:trivial}). In fact, it is clear from its definition
($\pi=\frac{TP}{TP+FP}$) that $\pi$ depends only on how the positives
($TP+FP$) are split between \emph{true} positives $TP$ and the
\emph{false} positives $FP$, and does not depend at all on the
cardinality of the positives. There is a breakup of ``symmetry''
between $\pi$ and $\rho$ here because, from the point of view of
classifier judgment (positives vs.\ negatives; this is the dichotomy
of interest in trivial acceptor vs.\ trivial rejector) the
``symmetric'' of $\rho$ ($\frac{TP}{TP+FN}$) is not $\pi$
($\frac{TP}{TP+FP}$) but \emph{c-precision}
($\pi^{c}=\frac{TN}{FP+TN}$), the ``contrapositive'' of $\pi$. In
fact, while $\rho$=1 and $\pi^{c}$=0 for the trivial acceptor,
$\pi^{c}$=1 and $\rho$=0 for the trivial rejector.}. Conversely, it is
well-known from everyday IR practice that higher levels of $\pi$ may
be obtained at the price of low values of $\rho$.

\begin{table}\begin{center}\begin{footnotesize}\begin{tabular}{|c
c||c|c|c|c|} \hline \multicolumn{2}{|c||}{\mbox{}} & Precision &
Recall & C-precision & C-recall \\
\multicolumn{2}{|c||}{\mbox{}} & \rule[-3ex]{0mm}{7ex}
$\displaystyle\frac{TP}{TP+FP}$ & $\displaystyle\frac{TP}{TP+FN}$ &
$\displaystyle\frac{TN}{FP+TN}$ & $\displaystyle\frac{TN}{TN+FN}$ \\
\hline \hline
\rule[-3ex]{0mm}{7ex} Trivial Rejector & TP=FP=0 & undefined &
$\displaystyle\frac{0}{FN}=0$ & $\displaystyle\frac{TN}{TN}=1$ &
$\displaystyle\frac{TN}{TN+FN}$ \\ \hline
\rule[-3ex]{0mm}{7ex}Trivial Acceptor & FN=TN=0 &
$\displaystyle\frac{TP}{TP+FP}$ & $\displaystyle\frac{TP}{TP}=1$ &
$\displaystyle\frac{0}{FP}=0$ & undefined \\ \hline
\rule[-3ex]{0mm}{7ex}Trivial ``Yes'' Collection & FP=TN=0 &
$\displaystyle\frac{TP}{TP}=1$ & $\displaystyle\frac{TP}{TP+FN}$ &
undefined & $\displaystyle\frac{0}{FN}=0$ \\ \hline
\rule[-3ex]{0mm}{7ex}Trivial ``No'' Collection & TP=FN=0 &
$\displaystyle\frac{0}{FP}=0$ & undefined &
$\displaystyle\frac{TN}{FP+TN}$ & $\displaystyle\frac{TN}{TN}=1$ \\
\hline
\end{tabular}\end{footnotesize}\caption{\label{tab:trivial}Trivial
cases in TC.}\end{center}\end{table}

In practice, by tuning $\tau_i$ a function $CSV_i:{\cal D}\rightarrow
\{T,F\}$ is tuned to be, in the words of Riloff and Lehnert
\citeyear{Riloff94}, more \emph{liberal} (i.e.\ improving $\rho_i$ to
the detriment of $\pi_i$) or more \emph{conservative} (improving
$\pi_i$ to the detriment of $\rho_i$)\footnote{While $\rho_i$ can
\emph{always} be increased at will by lowering $\tau_i$,
\emph{usually} at the cost of decreasing $\pi_i$, $\pi_i$ can {\em
usually} be increased at will by raising $\tau_i$, \emph{always} at
the cost of decreasing $\rho_i$. This kind of tuning is only possible
for $CSV_i$ functions with values in $[0,1]$; for binary-valued
$CSV_i$ functions tuning is not always possible, or is anyway more
difficult (see e.g.\ \cite[page 66]{Weiss99}).}. A classifier should
thus be evaluated by means of a measure which combines $\pi$ and
$\rho$\footnote{An exception is single-label TC, in which $\pi$ and
$\rho$ are not independent of each other: if a document $d_j$ has been
classified under a wrong category $c_{s}$ (thus decreasing $\pi_{s}$)
this also means that it has {\em not} been classified under the right
category $c_t$ (thus decreasing $\rho_{t}$). In this case either $\pi$
or $\rho$ can be used as a measure of effectiveness.}. Various such
measures have been proposed, among which the most frequent are:

\begin{enumerate}

\item \emph{11-point average precision}: threshold $\tau_i$ is
repeatedly tuned so as to allow $\rho_i$ to take up values of 0.0, .1,
\ldots, .9, 1.0; $\pi_i$ is computed for these 11 different values of
$\tau_i$, and averaged over the 11 resulting values. This is analogous
to the standard evaluation methodology for ranked IR systems, and may
be used

\begin{enumerate}

\item with categories in place of IR queries. This is most frequently
used for document-ranking classifiers (see e.g
\cite{Schutze95,Yang94a,Yang99a,Yang97});

\item with test documents in place of IR queries and categories in
place of documents. This is most frequently used for category-ranking
classifiers (see e.g.\ \cite{Lam99a,Larkey96,Schapire00,Wiener95}). In
this case if macroaveraging is used it needs to be redefined on a
per-document, rather than per-category basis.

\end{enumerate}

\noindent This measure does not make sense for binary-valued $CSV_i$
functions, since in this case $\rho_i$ may not be varied at will.

\item the \emph{breakeven} point, i.e.\ the value at which $\pi$
equals $\rho$ (e.g.
\cite{Apte94,Cohen99,Dagan97,Joachims98,Joachims99,Lewis92,Lewis94,Moulinier96a,Ng97,Yang99a}).
This is obtained by a process analogous to the one used for 11-point
average precision: a plot of $\pi$ as a function of $\rho$ is computed
by repeatedly varying the thresholds $\tau_i$; breakeven is the value
of $\rho$ (or $\pi$) for which the plot intersects the $\rho=\pi$
line. This idea relies on the fact that by decreasing the $\tau_i$'s
from 1 to 0, $\rho$ always increases monotonically from 0 to 1 and
$\pi$ usually decreases monotonically from a value near 1 to
$\frac{1}{|{\cal C}|}\sum_{i=1}^{|{\cal C}|}g_{Te}(c_i)$. If for no
values of the $\tau_i$'s $\pi$ and $\rho$ are exactly equal, the
$\tau_i$'s are set to the value for which $\pi$ and $\rho$ are
closest, and an \emph{interpolated breakeven} is computed as the
average of the values of $\pi$ and $\rho$\footnote{Breakeven, first
proposed by Lewis \citeyear{Lewis92,Lewis92a}, has been recently
criticized. Lewis himself (see his message of 11 Sep 1997 10:49:01 to
the DDLBETA text categorization mailing list -- quoted with permission
of the author) points out that breakeven is not a good effectiveness
measure, since (i) there may be no parameter setting that yields the
breakeven; in this case the final breakeven value, obtained by
interpolation, is artificial; (ii) to have $\rho$ equal $\pi$ is not
necessarily desirable, and it is not clear that a system that achieves
high breakeven can be tuned to score high on other effectiveness
measures. Yang \citeyear{Yang99a} also notes that when for no value of
the parameters $\pi$ and $\rho$ are close enough, interpolated
breakeven may not be a reliable indicator of effectiveness.}.

\item the \emph{$F_\beta$ function} \cite[Chapter 7]{vanRijsbergen79},
for some $0\leq \beta \leq +\infty$ (e.g.
\cite{Cohen95a,Cohen99,Lewis94a,Lewis95,Moulinier96,Ruiz99}), where
\begin{eqnarray}\nonumber
F_\beta=\displaystyle\frac{(\beta^2+1)\pi\rho}{\beta^2 \pi +
\rho}\end{eqnarray}

\noindent Here $\beta$ may be seen as the relative degree of
importance attributed to $\pi$ and $\rho$. If $\beta=0$ then $F_\beta$
coincides with $\pi$, whereas if $\beta=+\infty$ then $F_\beta$
coincides with $\rho$. Usually, a value $\beta=1$ is used, which
attributes equal importance to $\pi$ and $\rho$. As shown in
\cite{Moulinier96,Yang99a}, the breakeven of a classifier $\Phi$ is
always less or equal than its $F_1$ value.

\end{enumerate}

\noindent Once an effectiveness measure is chosen, a classifier can be
tuned (e.g.\ thresholds and other parameters can be set) so that the
resulting effectiveness is the best achievable by that classifier.
Tuning a parameter $p$ (be it a threshold or other) is normally done
experimentally. This means performing repeated experiments on the
validation set with the values of the other parameters $p_k$ fixed (at
a default value, in the case of a yet-to-be-tuned parameter $p_k$, or
at the chosen value, if the parameter $p_k$ has already been tuned)
and with different values for parameter $p$. The value that has
yielded the best effectiveness is chosen for $p$.


\subsection{Benchmarks for text categorization}
\label{sec:testcollections}

Standard \emph{benchmark collections} that can be used as initial
corpora for TC are publically available for experimental purposes. The
most widely used is the {\sf Reuters} collection, consisting of a set
of newswire stories classified under categories related to economics.
The {\sf Reuters} collection accounts for most of the experimental
work in TC so far. Unfortunately, this does not always translate into
reliable comparative results, in the sense that many of these
experiments have been carried out in subtly different conditions.

In general, different sets of experiments may be used for
cross-classifier comparison only if the experiments have been
performed

\begin{enumerate}

\item \label{item:caveatcollection}
on exactly the same collection (i.e.\ same documents and same
categories);

\item \label{item:caveatsplit} with the same ``split'' between
training set and test set;

\item \label{item:caveatmeasure} with the same evaluation measure and,
whenever this measure depends on some parameters (e.g.\ the utility
matrix chosen), with the same parameter values.

\end{enumerate}

\noindent Unfortunately, a lot of experimentation, both on {\sf
Reuters} and on other collections, has not been performed with these
\textit{caveat} in mind: by testing three different classifiers on
five popular versions of {\sf Reuters}, Yang \citeyear{Yang99a} has
shown that a lack of compliance with these three conditions may make
the experimental results hardly comparable among each other. Table
\ref{tab:comparativeresults} lists the results of all experiments
known to us performed on five major versions of the {\sf Reuters}
benchmark: {\sf Reuters-22173 ``ModLewis''} (column \#1), {\sf
Reuters-22173 ``ModApt\'e''} (column \#2), {\sf Reuters-22173
``ModWiener''} (column \#3), {\sf Reuters-21578 ``ModApt\'e''} (column
\#4) and {\sf Reuters-21578[10] ``ModApt\'e''} (column
\#5)\footnote{\label{foo:reuters} The {\sf Reuters-21578} collection
may be freely downloaded for experimentation purposes from {\tt
http://www.research.att.com/\~{}lewis/reuters21578.html} and is now
considered the ``standard'' variant of {\sf Reuters}. We do not cover
experiments performed on variants of {\sf Reuters} different from the
five listed because the small number of authors that have used the
same variant makes the reported results difficult to interpret. This
includes experiments performed on the original {\sf Reuters-22173
``ModHayes''} \cite{Hayes90a} and {\sf Reuters-21578 ``ModLewis''}
\cite{Cohen99}.}. Only experiments that have computed either a
breakeven or $F_1$ have been listed, since other less popular
effectiveness measures do not readily compare with these.

Note that only results belonging to the same column are directly
comparable. In particular, Yang \citeyear{Yang99a} showed that
experiments carried out on {\sf Reuters-22173 ``ModLewis''} (column
\#1) are not directly comparable with those using the other three
versions, since the former strangely includes a significant percentage
(58\%) of ``unlabelled'' test documents which, being negative examples
of all categories, tend to depress effectiveness. Also, experiments
performed on {\sf Reuters-21578[10] ``ModApt\'e''} (column \#5) are
not comparable with the others, since this collection is the
restriction of {\sf Reuters-21578 ``ModApt\'e''} to the 10 categories
with the highest generality, and is thus an obviously ``easier''
collection.

\begin{table}[t]\begin{center}\begin{tiny}\begin{tabular}{||c|c|c||c|c|c|c|c||}
\hline\hline
 & & & \textbf{\#1} & \textbf{\#2} & \textbf{\#3} & \textbf{\#4} &
 {\bf \#5} \\
 \hline\hline
 & & \textbf{\# of documents} & 21,450 & 14,347 & 13,272 & 12,902 &
 12,902 \\
 & & \textbf{\# of training documents} & 14,704 & 10,667 & 9,610 &
 9,603 & 9,603\\
 & & \textbf{\# of test documents} & 6,746 & 3,680 & 3,662 & 3,299 &
 3,299 \\
 & & \textbf{\# of categories} & 135 & 93 & 92 & 90 & 10 \\
 \hline\hline
 \textbf{System} & \textbf{Type} & \textbf{Results reported by} & & &
 & & \\
 \hline\hline
 \textsc{Word} & (non-learning) & \cite{Yang99a} & .150 & .310 & .290
 & & \\
 \hline
 & probabilistic & \cite{Dumais98} & & & & .752 & .815 \\
 & probabilistic & \cite{Joachims98} & & & & & .720 \\
 & probabilistic & \cite{Lam97} & .443 (M$F_1$) & & & & \\
 \textsc{PropBayes} & probabilistic & \cite{Lewis92} & .650 & & & & \\
 \textsc{Bim} & probabilistic & \cite{Li99} & & & & .747 & \\
 & probabilistic & \cite{Li99} & & & & .773 & \\
 \textsc{Nb} & probabilistic & \cite{Yang99} & & & & .795 & \\
 \hline
 & decision trees & \cite{Dumais98} & & & & & .884 \\
 C4.5 & decision trees & \cite{Joachims98} & & & & & .794 \\
 \textsc{Ind} & decision trees & \cite{Lewis94} & .670 & & & & \\
 \hline
 \textsc{Swap-1} & decision rules & \cite{Apte94} & & .805 & & & \\
 \textsc{Ripper} & decision rules & \cite{Cohen99} & .683 & .811 & &
 .820 & \\
 \textsc{SleepingExperts} & decision rules & \cite{Cohen99} &
 \textbf{.753} & .759 & & .827 & \\
 \textsc{Dl-Esc} & decision rules & \cite{Li99} & & & & .820 & \\
 \textsc{Charade} & decision rules & \cite{Moulinier96a} & &.738 & & &
 \\
 \textsc{Charade} & decision rules & \cite{Moulinier96} & & .783
 ($F_1$) & & & \\
 \hline
 \textsc{Llsf} & regression & \cite{Yang99a} & & .855 & .810 & & \\
 \textsc{Llsf} & regression & \cite{Yang99} & & & & .849 & \\
 \hline
 \textsc{BalancedWinnow} & on-line linear & \cite{Dagan97} & .747 (M)
 & .833 (M) & & & \\
 \textsc{Widrow-Hoff} & on-line linear & \cite{Lam98} & & & & .822 &
 \\
 \hline
 \textsc{Rocchio} & batch linear & \cite{Cohen99} & .660 & .748 & &
 .776 & \\
 \textsc{FindSim} & batch linear & \cite{Dumais98} & & & & .617 & .646
 \\
 \textsc{Rocchio} & batch linear & \cite{Joachims98} & & & & & .799 \\
 \textsc{Rocchio} & batch linear & \cite{Lam98} & & & & .781 & \\
 \textsc{Rocchio} & batch linear & \cite{Li99} & & & & .625 & \\
 \hline
 \textsc{Classi} & neural network & \cite{Ng97} & & .802 & & & \\
 \textsc{Nnet} & neural network & \cite{Yang99} & & & & .838 & \\
 & neural network & \cite{Wiener95} & & & \textbf{.820} & & \\
 \hline
 \textsc{Gis-W} & example-based & \cite{Lam98} & & & & .860 & \\
 k-NN & example-based & \cite{Joachims98} & & & & & .823 \\
 k-NN & example-based & \cite{Lam98} & & & & .820 & \\
 k-NN & example-based & \cite{Yang99a} & .690 & .852 & {\bf .820} & &
 \\
 k-NN & example-based & \cite{Yang99} & & & & .856 & \\
 \hline
 & SVM & \cite{Dumais98} & & & & .870 & \textbf{.920} \\
 \textsc{SvmLight} & SVM & \cite{Joachims98} & & & & & .864 \\
 \textsc{SvmLight} & SVM & \cite{Li99} & & & & .841 & \\
 \textsc{SvmLight} & SVM & \cite{Yang99} & & & & .859 & \\
 \hline
 \textsc{AdaBoost.MH} & committee & \cite{Schapire00} & & {\bf .860} &
 & & \\
 & committee & \cite{Weiss99} & & & & \textbf{.878} & \\
 \hline
 & Bayesian net & \cite{Dumais98} & & & & .800 & .850 \\
 & Bayesian net & \cite{Lam97} & .542 (M$F_1$) & & & & \\
 \hline \hline \end{tabular}
 \caption{\label{tab:comparativeresults}Comparative results among
 different classifiers obtained on five different version of {\sf
 Reuters}. Unless otherwise noted, entries indicate the microaveraged
 breakeven point; within parentheses, ``M'' indicates macroaveraging
 and ``$F_1$'' indicates use of the $F_1$ measure. {\bf Boldface}
 indicates the best performer on the collection.
 }\end{tiny}\end{center}\end{table}

Other test collections that have been frequently used are

\begin{itemize}

\item the {\sf OHSUMED} collection, set up by Hersh et al.
\citeyear{Hersh94} and used in
\cite{Joachims98,Lam98,Lam99a,Lewis96,Ruiz99,Yang97}\footnote{The {\sf
OHSUMED} collection may be freely downloaded for experimentation
purposes from {\tt ftp://medir.ohsu.edu/pub/ohsumed}}. The documents
are titles or title-plus-abstract's from medical journals ({\sf
OHSUMED} is actually a subset of the {\sf Medline} document base); the
categories are the ``postable terms'' of the MESH thesaurus.

\item the {\sf 20 Newsgroups} collection, set up by Lang
\citeyear{Lang95} and used in
\cite{Baker98,Joachims97,McCallum98,McCallum98b,Nigam00,Schapire00}.
The documents are messages posted to Usenet newsgroups, and the
categories are the newsgroups themselves.

\item the {\sf AP} collection, used in
\cite{Cohen95a,Cohen95,Cohen99,Lewis94c,Lewis94a,Lewis96,Schapire00,Schapire98}.

\end{itemize}

\noindent We will not cover the experiments performed on these
collections for the same reasons as those illustrated in Footnote
\ref{foo:reuters}, i.e.\ because in no case a significant enough
number of authors have used the same collection in the same
experimental conditions, thus making comparisons difficult.


\subsection{Which text classifier is best?} \label{sec:best}

The published experimental results, and especially those listed in
Table \ref{tab:comparativeresults}, allow us to attempt some
considerations on the comparative performance of the TC methods
discussed. However, we have to bear in mind that comparisons are
reliable only when based on experiments performed by the same author
under carefully controlled conditions. They are instead more
problematic when they involve different experiments performed by
different authors. In this case various ``background conditions'',
often extraneous to the learning algorithm itself, may influence the
results. These may include, among others, different choices in
pre-processing (stemming, etc.), indexing, dimensionality reduction,
classifier parameter values, etc., but also different standards of
compliance with safe scientific practice (such as tuning parameters on
the test set rather than on a separate validation set), which often
are not discussed in the published papers.

Two different methods may thus be applied for comparing classifiers
\cite{Yang99a}:

\begin{itemize}

\item \emph{direct comparison}: classifiers $\Phi'$ and $\Phi''$ may
be compared when they have been tested on the same collection
$\Omega$, usually by the same researchers and with the same background
conditions. This is the more reliable method.

\item \emph{indirect comparison}: classifiers $\Phi'$ and $\Phi''$ may
be compared when

\begin{enumerate}

\item \label{item:thetwo} they have been tested on collections
$\Omega'$ and $\Omega''$, respectively, typically by different
researchers and hence with possibly different background conditions;

\item \label{item:baseline} one or more ``baseline'' classifiers
$\overline{\Phi}_1, \ldots, \overline{\Phi}_m$ have been tested on
both $\Omega'$ and $\Omega''$ by the direct comparison method.

\end{enumerate}

Test \ref{item:baseline} gives an indication on the relative
``hardness'' of $\Omega'$ and $\Omega''$; using this and the results
from Test \ref{item:thetwo} we may obtain an indication on the
relative effectiveness of $\Phi'$ and $\Phi''$. For the reasons
discussed above, this method is less reliable.

\end{itemize}

\noindent A number of interesting conclusions can be drawn from Table
\ref{tab:comparativeresults} by using these two methods. Concerning
the relative ``hardness'' of the five collections, if by
$\Omega'>\Omega''$ we indicate that $\Omega'$ is a harder collection
that $\Omega''$, there seems to be enough evidence that {\sf
Reuters-22173 ``ModLewis''} $\gg$ {\sf Reuters-22173 ``ModWiener''}
$>$ {\sf Reuters-22173 ``ModApt\'e''} $\approx$ {\sf Reuters-21578
``ModApt\'e''} $>$ {\sf Reuters-21578[10] ``ModApt\'e''}. These facts
are unsurprising; in particular, the first and the last inequalities
are a direct consequence of the peculiar characteristics of {\sf
Reuters-22173 ``ModLewis''} and {\sf Reuters-21578[10] ``ModApt\'e''}
discussed in Section \ref{sec:testcollections}.

Concerning the relative performance of the classifiers, remembering
the considerations above we may attempt a few conclusions:

\begin{itemize}

\item Boosting-based classifier committees, support vector machines,
example-based methods, and regression methods deliver top-notch
performance. There seems to be no sufficient evidence to decidedly opt
for either method; efficiency considerations or application-dependent
issues might play a role in breaking the tie.

\item Neural networks and on-line linear classifiers work very well,
although slightly worse than the previously mentioned methods.

\item Batch linear classifiers (Rocchio) and probabilistic Na\"{\i}ve
Bayes classifiers look the worst of the learning-based classifiers.
For Rocchio, these results confirm earlier results by Sch\"utze et al.
\citeyear{Schutze95}, who had found three classifiers based on linear
discriminant analysis, linear regression, and neural networks, to
perform about 15\% better than Rocchio. However, recent results by
Schapire et al.\ \citeyear{Schapire98} rank Rocchio along the best
performers once near-positives are used in training.

\item The data in Table \ref{tab:comparativeresults} are hardly
sufficient to say anything about decision trees. However, the work by
Dumais et al.\ \citeyear{Dumais98} in which a decision tree classifier
was shown to perform nearly as well as their top performing system (a
SVM classifier) will probably renew the interest in decision trees, an
interest that had dwindled after the unimpressive results reported in
earlier literature \cite{Cohen99,Joachims98,Lewis94c,Lewis94}.
 
\item By far the lowest performance is displayed by {\sc Word}, a
classifier implemented by Yang \citeyear{Yang99a} and not including
any learning component\footnote{\textsc{Word} is based on the
comparison between documents and category names, each treated as a
vector of weighted terms in the vector space model. {\sc Word} was
implemented by Yang with the only purpose of determining the
difference in effectiveness that adding a learning component to a
classifier brings about. \textsc{Word} is actually called STR in
\cite{Yang94a,Yang94}. Another no-learning classifier is proposed in
\cite{Wong96}.}.

\end{itemize}

\noindent Concerning \textsc{Word} and no-learning classifiers, for
completeness we should recall that one of the highest effectiveness
values reported in the literature for the {\sf Reuters} collection (a
.90 breakeven) belongs to \textsc{Construe}, a manually constructed
classifier. However, this classifier has never been tested on the
\emph{standard} variants of {\sf Reuters} mentioned in Table
\ref{tab:comparativeresults}, and it is not clear \cite{Yang99a}
whether the (small) test set of {\sf Reuters-22173 ``ModHayes''} on
which the .90 breakeven value was obtained was chosen randomly, as
safe scientific practice would demand. Therefore, the fact that this
figure is indicative of the performance of {\sc Construe}, and of the
manual approach it represents, has been convincingly questioned
\cite{Yang99a}.

It is important to bear in mind that the considerations above are not
absolute statements (if there may be any) on the comparative
effectiveness of these TC methods. One of the reasons is that a
particular applicative context may exhibit very different
characteristics from the ones to be found in {\sf Reuters}, and
different classifiers may respond differently to these
characteristics. An experimental study by Joachims
\citeyear{Joachims98} involving support vector machines, $k$-NN,
decision trees, Rocchio and Na\"{\i}ve Bayes, showed all these
classifiers to have similar effectiveness on categories with $\geq
300$ positive training examples each. The fact that this experiment
involved the methods which have scored best (support vector machines,
$k$-NN) and worst (Rocchio and Na\"{\i}ve Bayes) according to Table
\ref{tab:comparativeresults} shows that applicative contexts different
from {\sf Reuters} may well invalidate conclusions drawn on this
latter.

Finally, a note is worth about statistical significance testing. Few
authors have gone to the trouble of validating their results by means
of such tests. These tests are useful for verifying how strongly the
experimental results support the claim that a given system $\Phi'$ is
better than another system $\Phi''$, or for verifying how much a
difference in the experimental setup affects the measured
effectiveness of a system $\Phi$. Hull \citeyear{Hull94} and Sch\"utze
et al.\ \citeyear{Schutze95} have been among the first to work in this
direction, validating their results by means of the \textsc{Anova}
test and the Friedman test; the former is aimed at determining the
significance of the difference in effectiveness between two methods in
terms of the ratio between this difference and the effectiveness
variability across categories, while the latter conducts a similar
test by using instead the rank positions of each method within a
category. Yang and Liu \citeyear{Yang99} define a full suite of
significance tests, some of which apply to microaveraged and some to
macroaveraged effectiveness. They apply them systematically to the
comparison between five different classifiers, and are thus able to
infer fine-grained conclusions about their relative effectiveness. For
other examples of significance testing in TC see
\cite{Cohen95a,Cohen95,Cohen98,Joachims97,Koller97,Lewis96,Wiener95}.


\section{Conclusion} \label{sec:conclusion}

Automated TC is now a major research area within the information
systems discipline, thanks to a number of factors

\begin{itemize}

\item Its domains of application are numerous and important, and given
the proliferation of documents in digital form they are bound to
increase dramatically in both number and importance.

\item It is indispensable in many applications in which the sheer
number of the documents to be classified and the short response time
required by the application make the manual alternative implausible.

\item It can improve the productivity of human classifiers in
applications in which no classification decision can be taken without
a final human judgment \cite{Larkey96}, by providing tools that
quickly ``suggest'' plausible decisions.

\item It has reached effectiveness levels comparable to those of
trained professionals. The effectiveness of manual TC is not 100\%
anyway \cite{Cleverdon84} and, more importantly, it is unlikely to be
improved substantially by the progress of research. The levels of
effectiveness of automated TC are instead growing at a steady pace,
and even if they will likely reach a plateau well below the 100\%
level, this plateau will probably be higher that the effectiveness
levels of manual TC.

\end{itemize}

\noindent One of the reasons why from the early '90s the effectiveness
of text classifiers has dramatically improved, is the arrival in the
TC arena of ML methods that are backed by strong theoretical
motivations. Examples of these are multiplicative weight updating
(e.g.\ the \textsc{Winnow} family, \textsc{Widrow-Hoff}, etc.),
adaptive resampling (e.g.\ boosting) and support vector machines,
which provide a sharp contrast with relatively unsophisticated and
weak methods such as Rocchio. In TC, ML researchers have found a
challenging application, since datasets consisting of hundreds of
thousands of documents and characterized by tens of thousands of terms
are widely available. This means that TC is a good benchmark for
checking whether a given learning technique can scale up to
substantial sizes. In turn, this probably means that the active
involvement of the ML community in TC is bound to grow.

The success story of automated TC is also going to encourage an
extension of its methods and techniques to neighbouring fields of
application. Techniques typical of automated TC have already been
extended successfully to the categorization of documents expressed in
slightly different media; for instance:

\begin{itemize}

\item very noisy text resulting from optical character recognition
\cite{Ittner95,Junker98}. In their experiments Ittner et al.
\citeyear{Ittner95} have found that, by employing noisy texts also in
the training phase (i.e.\ texts affected by the same source of noise
that is also at work in the test documents), effectiveness levels
comparable to those obtainable in the case of standard text can be
achieved.

\item speech transcripts \cite{Myers00,Schapire00}. For instance,
Schapire and Singer \citeyear{Schapire00} classify answers given to a
phone operator's request ``{\tt How may I help you?}", so as to be
able to route the call to a specialized operator according to call
type.

\end{itemize}

\noindent Concerning other more radically different media, the
situation is not as bright (however, see \cite{Lim99} for an
interesting attempt at image categorization based on a textual
metaphor). The reason for this is that capturing real semantic content
of non-textual media by automatic indexing is still an open problem.
While there are systems that attempt to detect content e.g.\ in images
by recognising shapes, colour distributions and texture, the general
problem of image semantics is still unsolved. The main reason is that
natural language, the language of the text medium, admits far fewer
variations than the ``languages'' employed by the other media. For
instance, while the concept of a house can be ``triggered'' by
relatively few natural language expressions such as {\tt house}, {\tt
houses}, {\tt home}, {\tt housing}, {\tt inhabiting}, etc., it can be
triggered by \emph{far more} images: the images of all the different
houses that exist, of all possible colours and shapes, viewed from all
possible perspectives, from all possible distances, etc. If we had
solved the multimedia indexing problem in a satisfactory way, the
general methodology that we have discussed in this paper for text
would also apply to automated multimedia categorization, and there are
reasons to believe that the effectiveness levels could be as high.
This only adds to the common sentiment that more research in automated
content-based indexing for multimedia documents is needed.


\section*{Acknowledgements} This paper owes a lot to the suggestions
and constructive criticism of Norbert Fuhr and David Lewis. Thanks
also to Umberto Straccia for comments on an earlier draft and to
Alessandro Sperduti for many fruitful discussions.


\bibliography{ATCbibliography,ATCbibliographyGrey,fab,fabspublications}
\bibliographystyle{esub2acm}

\end{document}